\documentclass[superscriptaddress,twocolumn]{revtex4}
\usepackage{times}
\usepackage[T1]{fontenc}
\usepackage[latin1]{inputenc}
\usepackage{verbatim}
\usepackage{amsmath}
\usepackage{graphicx}
\usepackage{amssymb}

\makeatletter

\providecommand{\tabularnewline}{\\}

\usepackage[english]{babel}
\makeatother
\begin{document}

\title{Constraints on transmission, dispersion, and density of states \\
in dielectric multilayers and stepwise potential barriers with arbitrary
layer arrangement}

\author{S.~V.~Zhukovsky}

\email{sergei@th.physik.uni-bonn.de}

\affiliation{Institute of Molecular and Atomic Physics, National Academy of Belarus,
Nezavisimosti~Ave. 70, 220072 Minsk, Belarus}

\affiliation{Physilakisches Institut, Universit\"at Bonn, Nussallee 12, D-53115
Bonn, Germany}

\author{S.~V.~Gaponenko}

\email{s.gaponenko@dragon.bas-net.by}

\affiliation{Institute of Molecular and Atomic Physics, National Academy of Belarus,
Nezavisimosti~Ave. 70, 220072 Minsk, Belarus}

\begin{abstract}
Normal-incidence transmission and dispersion properties of optical
multilayers and one-dimensional stepwise potential barriers in the
non-tunneling regime are analytically investigated. The optical paths
of every constituent layer in a multilayer structure, as well as the
parameters of every step of the stepwise potential barrier, are constrained
by a generalized quarter-wave condition. No other restrictions on
the structure geometry is imposed, i.e., the layers are arranged arbitrarily.
We show that the density of states (DOS) spectra of the multilayer
or barrier in question are subject to integral conservation rules
similar to the Barnett-Loudon sum rule but ocurring within a finite
frequency or energy interval. In the optical case, these frequency
intervals are regular. For the potential barriers, only non-periodic
energy intervals can be present in the spectrum of any given structure,
and only if the parameters of constituent potential steps are properly
chosen.

The integral conservation relations derived analytically have also
been verified numerically. The relations can be used in dispersion-engineered
multilayer-based devices, e.g., ultrashort pulse compressors or ultracompact
optical delay lines, as well as to design multiple-quantum-well electronic
heterostructures with engineered DOS.
\end{abstract}
\maketitle

\section{Introduction\label{sec:introduction}}

Over the centuries, the concept of homogeneity has played a major
part in both mathematics and physics. The very name of a fundamental
monograph on electrodynamics \cite{LandauLifshits}, \emph{Electrodynamics
of Continuous Media}, suggests that there should also exist electrodynamics
of \emph{discontinuous} media, quite distinct and yet unexplored.
Indeed, most real-world physical phenomena and processes are usually
neither continuous nor homogeneous, and all seemingly homogeneous
substances are in fact discontinuous on the molecular and atomic level.
The reason why the concept of homogeneous media is applicable and
produces good results in electrodynamics is that, in the first place,
the microscopic structure is so much smaller than typical electromagnetic
wavelengths that an effective-medium approximation is valid. Secondly,
many macroscopic systems can be broken up into several homogeneous
parts, the relatively large size of which making the studies of the
whole system comparatively simple.

The intermediate case of \emph{mesoscopic structures} where inhomogeneities
appear on the scale not minuscule enough to use an effective-medium
approach but not too large to allow finite-size effects to be neglected
has appeared more or less recently. This was largely motivated by
the advancement of technology, allowing such structures to be fabricated
and characterized. Even the first steps in this direction have already
caused major advancements. The onset of semiconductor heterostructures
was a breakthrough in electronics, the pioneers in the area awarded
the Nobel Prize in 2000 \cite{NobelReview}. The introduction of quantum
mesoscopic systems such as nanocrystals and quantum dots opens new
horizons in many areas, including biological sensor design and solid-state
quantum computation (see, e.g., \cite{SVBook} and references therein).
The introduction of micro- and nanostructured optical materials has
opened up whole new areas of photonic crystal research, integrated
optics, and the newly-emerging metamaterial physics (see \cite{PCBooks}),
with innumerable applications in telecommunication. 

It appears that by arranging the matter in a mesoscopically structured
fashion, one can engineer its properties (e.g., electronic and/or
optical) with considerable freedom. One can achieve as rich a variety
as seen among natural substances due to a known diversity in their
molecular-sized chemical composition. This freedom is especially increased
when the concept of structuring is extended beyond periodicity (see
the recent review \cite{MaciaReview}), such as in quasiperiodic \cite{AllQP}
or fractal media \cite{AllFractals}. Often it is even possible to
design a structured medium in order to achieve the chosen desired
properties \cite{Optimization,Subhashish}. Structured media can even
exhibit optical properties beyond what occurs in natural materials,
e.g., negative refraction \cite{AllNegative} and the ability to slow
down or stop light pulses \cite{AllSlowLight}. 

All physical phenomena that involve interaction between light and
matter appear to be altered in inhomogeneous media. This alteration
is believed to be a fundamental physical principle involving modification
of the properties of the vacuum (electromagnetic or electronic) in
the vicinity of inhomogeneities. Such modification is generally described
using the concept of the density of states (DOS) \cite{SVRaman};
however, this concept is not without discussion points as regards
definition of the DOS in finite vs. infinite media \cite{Busch2D}.
Despite those difficulties, the DOS concept appears a promising candidate
for a universal approach towards consistent description of physical
processes in arbitrarily inhomogeneous media.

As an example, it should be clear that the above mentioned modification
of the vacuum cannot be totally arbitrary. Causality had been shown
to restrict the modification of spontaneous emission rate by spectral
redistribution with the total emission rate remaining unaffected (the
Barnett-Loudon sum rule \cite{BarnettLoudon}). Since spontaneous
emission is related to the DOS, this rule can be expressed as impossibility
to change the total {}``number'' of states, but only to redistribute
them spectrally, which appears to be intuitively clear and heuristically
potent.

In this paper we report on another, related limitation concerning
the modification of transmission and dispersion properties of optical
dielectric multilayers as well as of electronic heterostructures consisting
of stepwise potential wells and barriers in the non-tunneling regime. 

For the optical case, making all constituent layers commensurate in
optical path produces a set of equidistant single-layer reflection-free
(SLRF) points $2m\omega_{0}$ where (and only where) the dispersion
relation of the structure coincides with that of a homogeneous medium
(i.e., $k(2m\omega_{0})\propto2m\omega_{0}$). We have found that
the optical DOS integrated between these points does not depend on
the structure geometry and does not change if the constituent layers
(whose optical paths are commensurate with respect to each other)
are rearranged. The degree of modification to the optical properties
as due to inhomogeneity of the structure is thus shown to be limited
not only in its amount but also in its extent. This means that not
only is a DOS enhancement in one spectral region compensated for in
some other region, but also the compensation must occur within the
distance $2\omega_{0}$ between the SLRF points, which is a spectral
interval preset by the structural parameters of the constituent element.

For the stepwise potential, similar relations have been shown to exist.
The single-layer reflection-free points do occur but are no longer
equidistant. For any given values of parameters for constituent elements,
there can be either none or a multitude of aperiodically located SLRF
points for all structures. In the latter case, the integral constraints
on the DOS can still be obtained, but they are more complex. In both
optical and quantum case, the analytical relations obtained have been
confirmed in numerical calculations. 

The paper is organized as follows. In Sec.~\ref{sec:grounds} we
introduce the reader to the structures under study and provide the
necessary basic notation. In Sec.~\ref{sec:dos}, we discuss the
concept of the DOS and its relation to the spectral properties of
the structure. In Secs.~\ref{sec:optical} and~\ref{sec:quantum}
we derive the constraints on the DOS for optical multilayers and for
binary stepwise potential barriers, respectively. In Sec.~\ref{sec:discussion}
the results obtained in the previous sections are compared and discussed.
Finally, Sec.~\ref{sec:summary} summarizes the paper.

\section{Optical and electronic heterostructures\label{sec:grounds}}

\begin{figure*}
\hfill{}\includegraphics[width=0.3\textwidth]{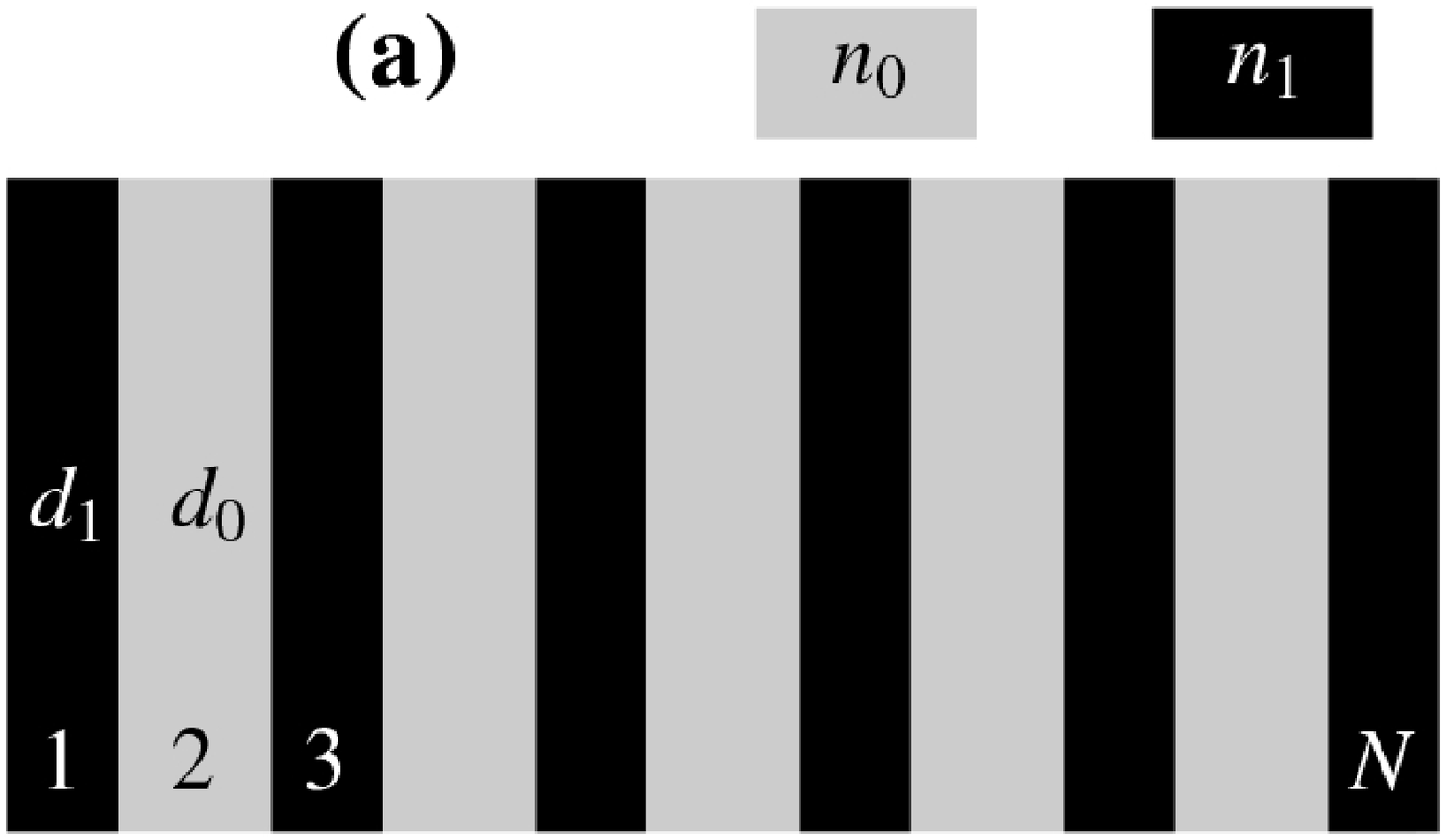}\hfill{}\includegraphics[width=0.3\textwidth]{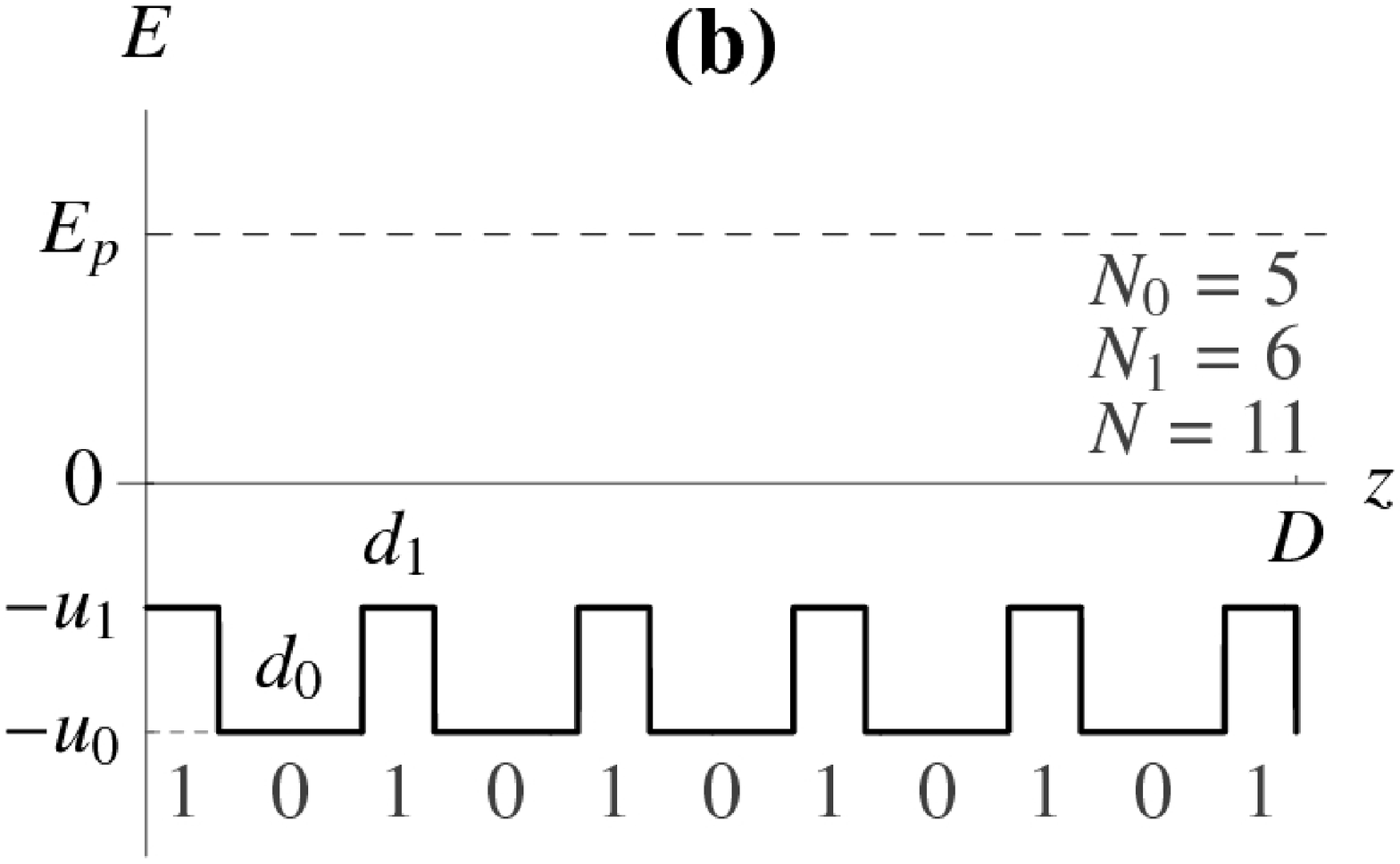}\hfill{}\includegraphics[width=0.3\textwidth]{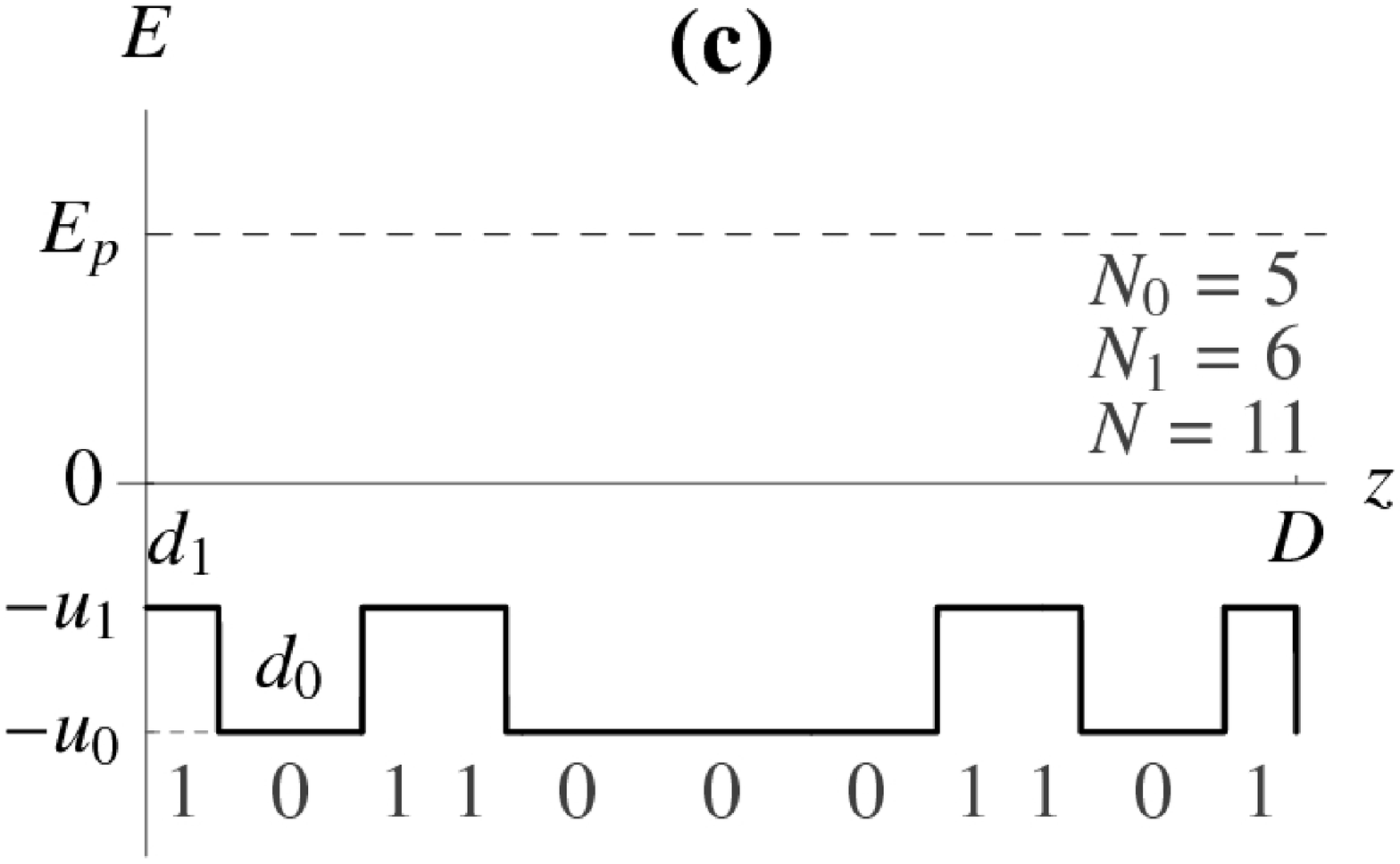}\hfill{}

\caption{A binary optical multilayer \textbf{(a)} together with its quantum
mechanical counterpart: a particle with energy~$E_{p}=\hbar\omega$
in 1D periodic \textbf{(b)} and non-periodic \textbf{(c)} binary stepwise
potential. \label{fig:structure}}
\end{figure*}

We start by considering a one-dimensional dielectric multilayer nanostructure
of $N$~layers, each layer having a thickness~$d_{j}$ and a refractive
index~$n_{j}$, infinite in the transverse directions and surrounded
on both sides by free space ($n_{0}=1$). Consider a normally incident
plane monochromatic wave propagating through such a structure. This
problem is one-dimensional, and unless the multilayer structure contains
optically anisotropic materials, it can be described using scalar
electric field governed by the scalar Helmholtz equation \cite{PCBooks}\begin{equation}
\frac{\partial^{2}}{\partial x^{2}}E(x)+\varepsilon(x)\frac{\omega^{2}}{c^{2}}E(x)=0.\label{eq:helmholtz}\end{equation}

Let $R(\omega)$ and $T(\omega)$ denote the complex (i.e., taking
into account the phase shift) reflection and transmission coefficients
of the multilayer structure, respectively. Let us now assume that
all the layers have such parameters that the optical path $n_{j}d_{j}$
is the same for any~$j$, so that\begin{equation}
n_{1}d_{1}=n_{2}d_{2}=\cdots=n_{j}d_{j}=\cdots=n_{N}d_{N}\equiv\pi c/2\omega_{0}.\label{eq:qwave}\end{equation}
where~$\omega_{0}$ is defined as the \emph{central frequency}. We
call any structure that conforms to Eq.~\eqref{eq:qwave} the \emph{quarter
wave} (QW) multilayer structure. 

It can be shown that for any even multiple of~$\omega_{0}$ the propagating
wave passes each constituent layer without reflection (no internal
reflections at the layer interfaces), and thus gains the phase shift
$\Delta\varphi=\frac{\omega}{c}n_{j}d_{j}$, which is the same for
all layers in view of the QW condition as expressed with Eq.~\eqref{eq:qwave}.
(See also \cite{ESPhaseShift} for more detail on phase relations
in Fresnel reflection from one layer.) As a result, the structure
becomes fully transparent ($\left|T(2m\omega_{0})\right|=1$) regardless
of the number or arrangement of constituent layers, and the total
phase shift becomes a simple sum of the shifts for all the layers:\begin{equation}
T(\omega_{m}=2m\omega_{0})=\exp\left[\mathrm{i}\frac{\omega}{c}\sum_{j=1}^{N}n_{j}d_{j}\right]=\exp\left(\mathrm{i}Nm\pi\right).\label{eq:fixed_phase}\end{equation}

Eq.~\eqref{eq:fixed_phase} essentially provides a set of equidistant
frequency points {[}we will call them \emph{single-layer reflection-free}
(SLRF) points] where the propagation phase (and hence, the wave number)
is linearly dependent on frequency regardless of the structure. Indeed,
the dispersion relation at these points ($D=\sum_{j=1}^{N}d_{j}$)
\begin{equation}
k_{m}(\omega_{m})=k(2m\omega_{0})=\frac{Nm\pi}{D}=\frac{N\lambda_{0}}{2D}\frac{\omega_{m}}{c}\label{eq:fixed_dispersion}\end{equation}
linearly relates $k_{m}$~and~$\omega_{m}$, as is the case for
a homogeneous medium. This linear dependence occurs only at the set
of SLRF points $\omega_{m}=2m\omega_{0}$, and Eqs.~\eqref{eq:fixed_phase}--\eqref{eq:fixed_dispersion}
are not true anywhere between these points. Note that among all the
transmission resonances present in a given multilayer's spectrum,
the SLRF points represent stationary waves without any correlations
on a length scale greater than the optical path of one constituent
layer. As such, they are naturally the least localized non-evanescent
eigenstates possible in any given QW multilayer.

Moreover, QW structures are known to possess spectral periodicity
in transmittance \cite{Nano03}\begin{equation}
\left|T(\omega+2m\omega_{0})\right|=\left|T(\omega)\right|\label{eq:qw_period}\end{equation}
and mirror symmetry within each period \cite{Nano03} \begin{equation}
\begin{array}{c}
\left|T\left((2m+1)\omega_{0}+\omega\right)\right|=\left|T\left((2m+1)\omega_{0}-\omega\right)\right|,\\
0<\omega<\omega_{0}.\end{array}\label{eq:qw_mirror}\end{equation}

Now let us note that Eq.~\eqref{eq:helmholtz} is isomorphic to the
Schr\"odinger equation governing a quantum particle with mass~$m_{p}$
and energy~$E_{p}$ in a stepwise potential~$u(x)$\begin{equation}
\frac{\hbar^{2}}{2m_{p}}\frac{\partial^{2}}{\partial x^{2}}\psi(x)+\left[E_{p}-u(x)\right]\psi(x)=0\label{eq:schroedinger}\end{equation}

If the potential function is a constant ($u(x)=-u_{0}$), the solution
of Eq.~\eqref{eq:schroedinger} is a plane-wave function \begin{equation}
\psi_{0}(x)=\exp\mathrm{i}kx=\exp\left[\mathrm{i}\left(\frac{\sqrt{2m_{p}}}{\hbar}\sqrt{E_{p}+u_{0}}\right)x\right],\label{eq:shr_planewave}\end{equation}
which is analogous to a plane optical wave with the wave vector $k=\left(\sqrt{2m_{p}}/\hbar\right)\sqrt{E_{p}+u_{0}}$.
If $E_{p}>u(x)$ for any~$x$, then $k$~is real and the particle
interacts with a potential barrier in the non-tunneling (Ramsauer)
regime. This is a quantum mechanical analogy to electromagnetic wave
propagation in a dielectric structure. Similar to the optical case,
one can consider a stepwise potential barrier consisting of~$N$
{}``elementary wells'' (Fig.~\ref{fig:structure}b,c). The role
of refractive index is taken by the potential energy~$u_{j}$ in
every step of the whole potential function. The frequency is replaced
by the particle energy~$E_{p}$, which can be expressed in terms
of de~Broglie frequency \cite{Analogy} as $E_{p}=\hbar\omega$.
The case $E_{p}<u(x)$, which causes imaginary wave vector in Eq.~\eqref{eq:shr_planewave},
is the tunneling case analogous to light propagation in absorbing
media (e.g., metals). It is outside the scope of the present paper.
To assure $E_{p}>u(x)$, let us assume $E_{p}\ge0$ and $u(x)<0$
from now on.

Furthermore, it is commonly known that if the potential represents
a single step ($u(x)=-u_{0}$ for, say, $x<x_{0}$ and $u(x)=-u_{1}$
otherwise), one can introduce the coefficients\begin{equation}
r_{01}=\frac{k_{0}-k_{1}}{k_{0}+k_{1}},\quad t_{01}=\frac{2k_{0}}{k_{0}+k_{1}},\label{eq:qm_barrier}\end{equation}
which, when squared, denote the probability of finding the impinging
quantum particle reflected or transmitted, respectively \cite{Analogy}.
One can name them the reflection and transmission coefficient for
matter waves, a potential step corresponding to a single interface
between dielectric media. Since the expressions for $r$~and~$t$
are the same (the wave vector~$k$ taking the role of refractive
index), one can use the same transfer-matrix formalism for determining
both the stationary electromagnetic wave distribution in a multilayer
\cite{YarivYeh} and the steady-state wave function for quantum particles
travelling through a complex stepwise potential \cite{SoukoulisNew}.
Note that Eq.~\eqref{eq:shr_planewave} indicates that the {}``refractive
index'' introduced in this way possesses an inherent quadratic dispersion.

Finally, we call a multilayer structure \emph{binary} if it can be
represented as consisting of two types of constituent layers (denoted
as binary digits $0$~and~$1$, following the notation in \cite{OptComm}),
to which two combinations of refractive index and thickness $(n_{0};d_{0})$
and $(n_{1};d_{1})$ are attributed. By arranging the $0$'s and $1$'s
in different sequences, it is possible to vary the geometry of the
structure very widely, making it periodic (if $0$~and~$1$ alternate,
as in $10101010101$), disordered (if the sequence is randomly determined),
or deterministically aperiodic (e.g., quasiperiodic \cite{AllQP}
or fractal \cite{AllFractals}). A binary potential barrier, with
constituent elementary wells associated with $(u_{0};d_{0})$ and
$(u_{1};d_{1})$, can be introduced likewise. For brevity, we will
occasionally use the term {}``layers'' for both types of constituent
elements.

Note that whenever the sequence contains two identical layers (e.g.,
{}``$00$'' in $1010100101$) , it will of course mean in practice
that the corresponding structure will contain a single layer with
thickness~$2d_{0}$. However, for the purpose of this work we will
regard such combinations as two separate constituent layers. The number
of layers of both types $N_{0}$~and~$N_{1}$, as well as their
total number $N=N_{0}+N_{1}$, will then remain the same regardless
of layer rearrangement, indicative of the transmission coefficient
phase at the SLRF points {[}see Eq.~\eqref{eq:fixed_phase}].

\section{Optical and quantum density of states\label{sec:dos}}

As mentioned in Section~\ref{sec:introduction}, any inhomogeneity
present in space is known to modify the properties of the quantum
or electromagnetic vacuum in its vicinity. This modification takes
the form of the change in the local DOS. It is believed to affect
all phenomena that involve light-matter interaction, such as spontaneous
emission or Raman scattering (see, e.g., \cite{SVRaman} and references
therein). 

\begin{figure*}
\includegraphics[width=0.33\textwidth]{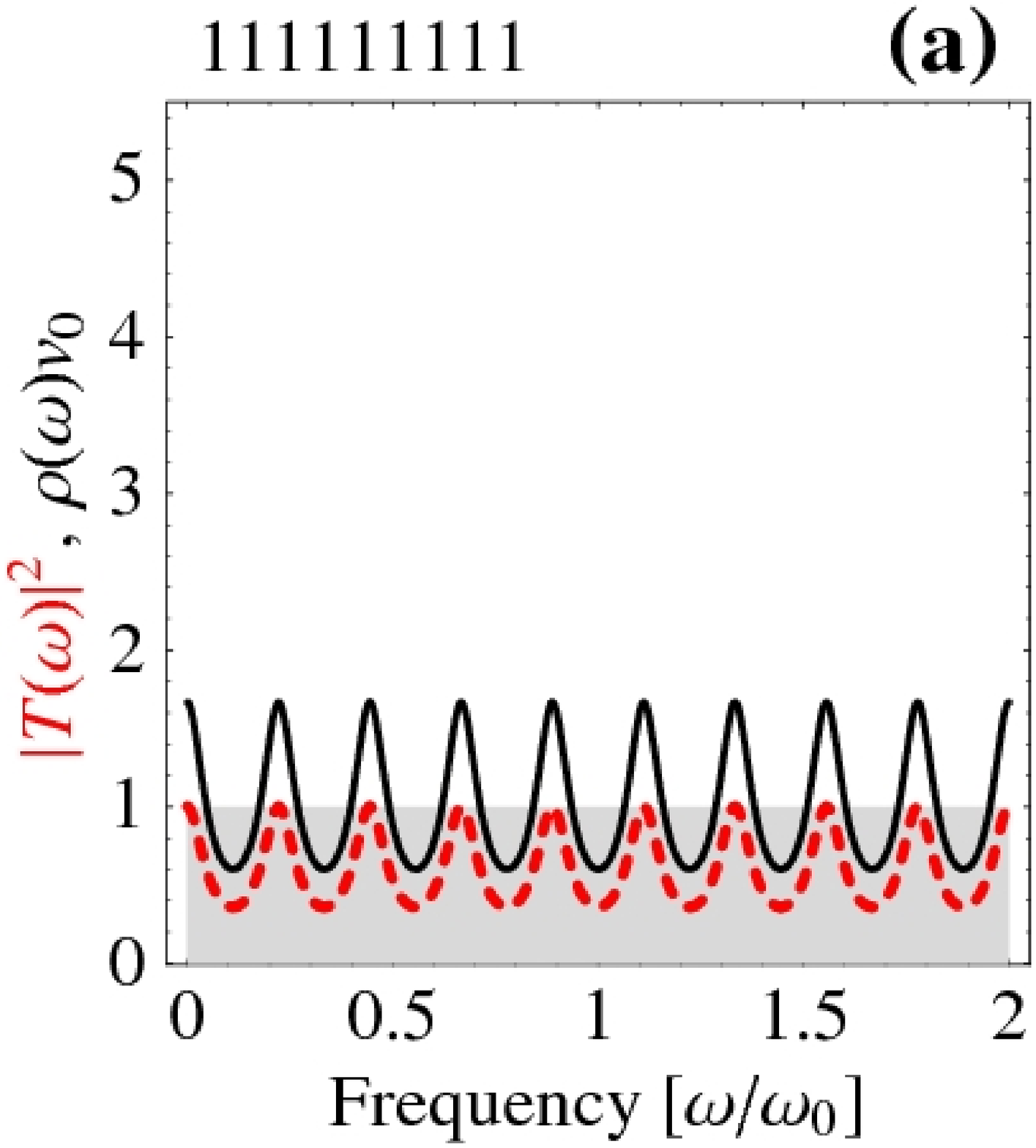}\includegraphics[width=0.33\textwidth]{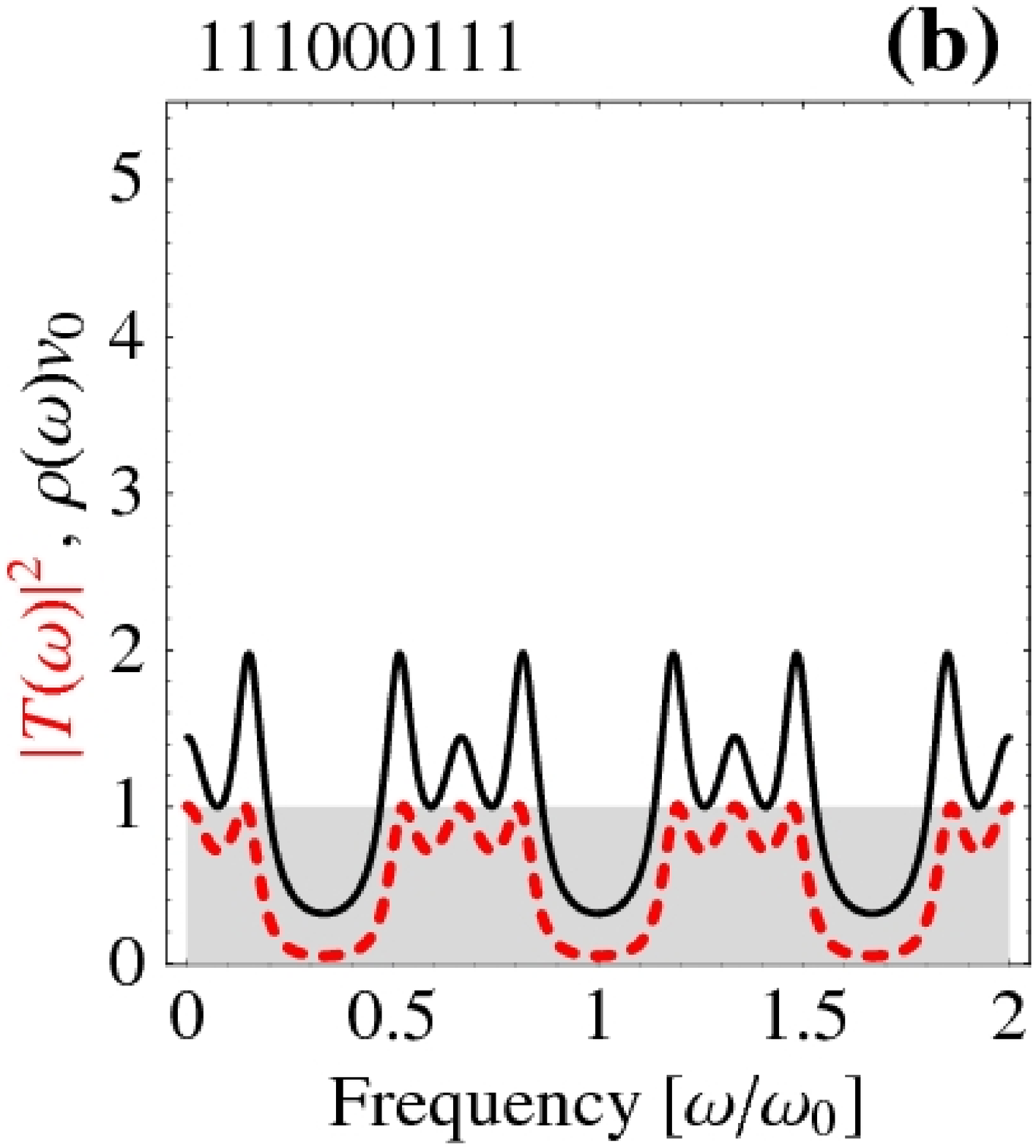}\includegraphics[width=0.33\textwidth]{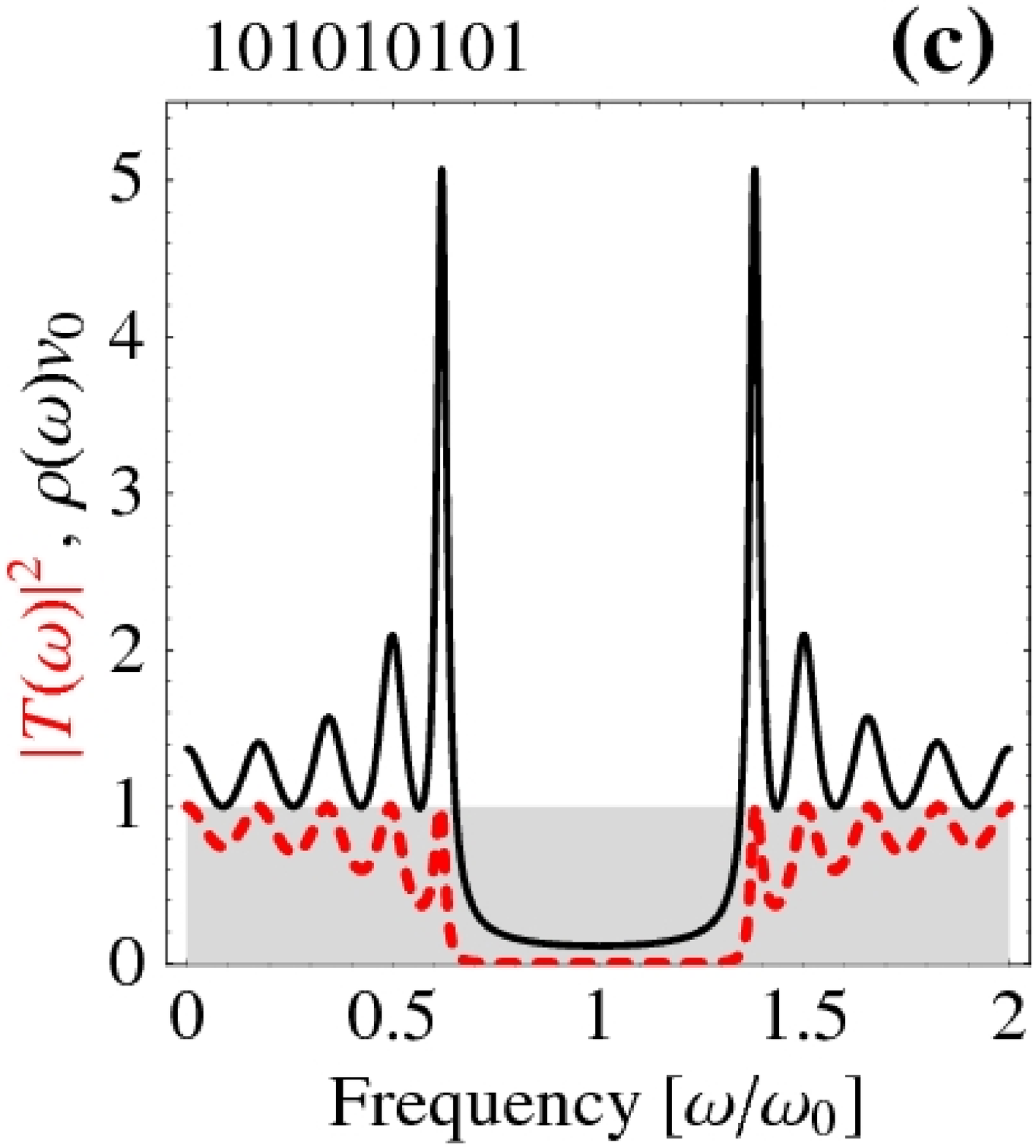}\medskip{}

\includegraphics[width=0.33\textwidth]{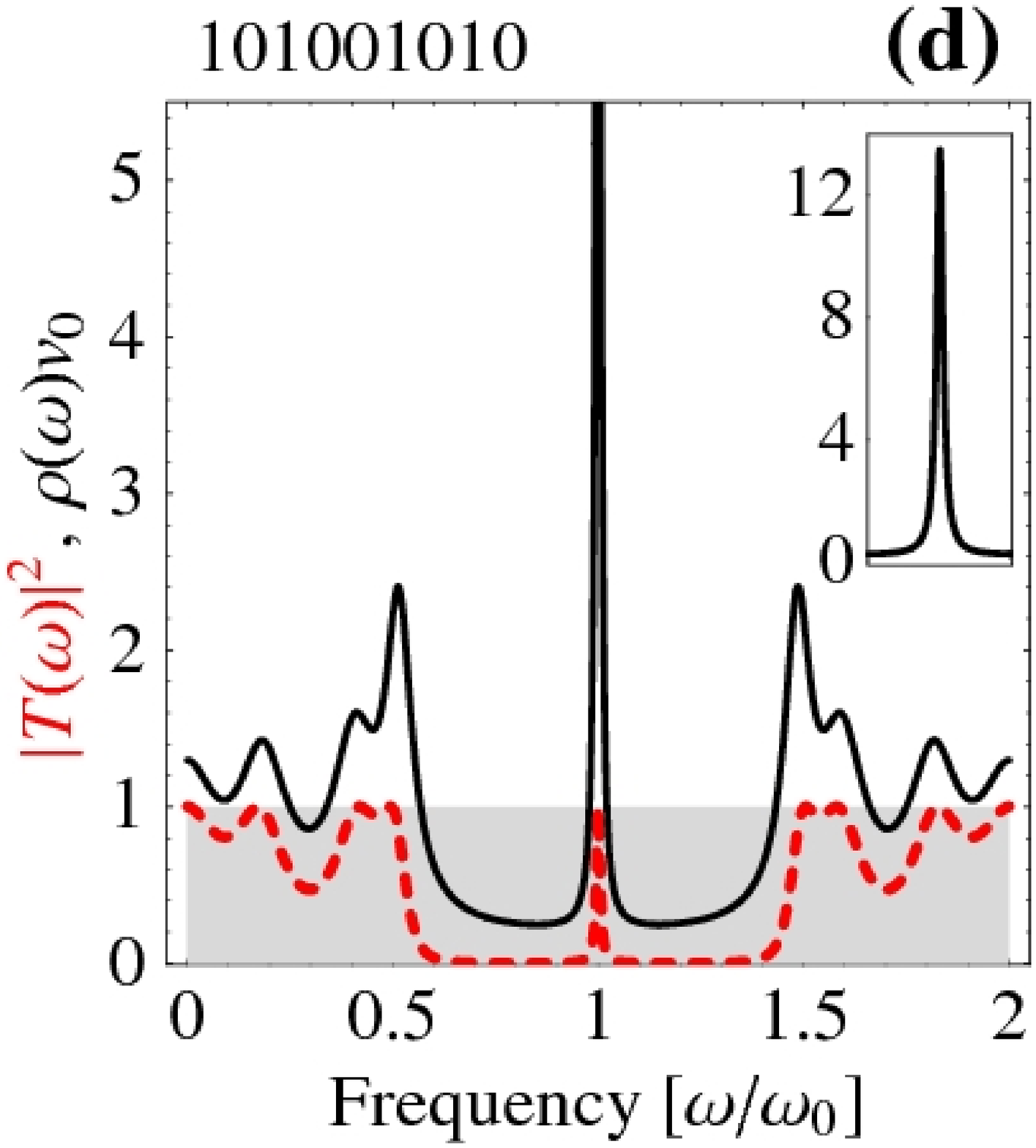}\includegraphics[width=0.33\textwidth]{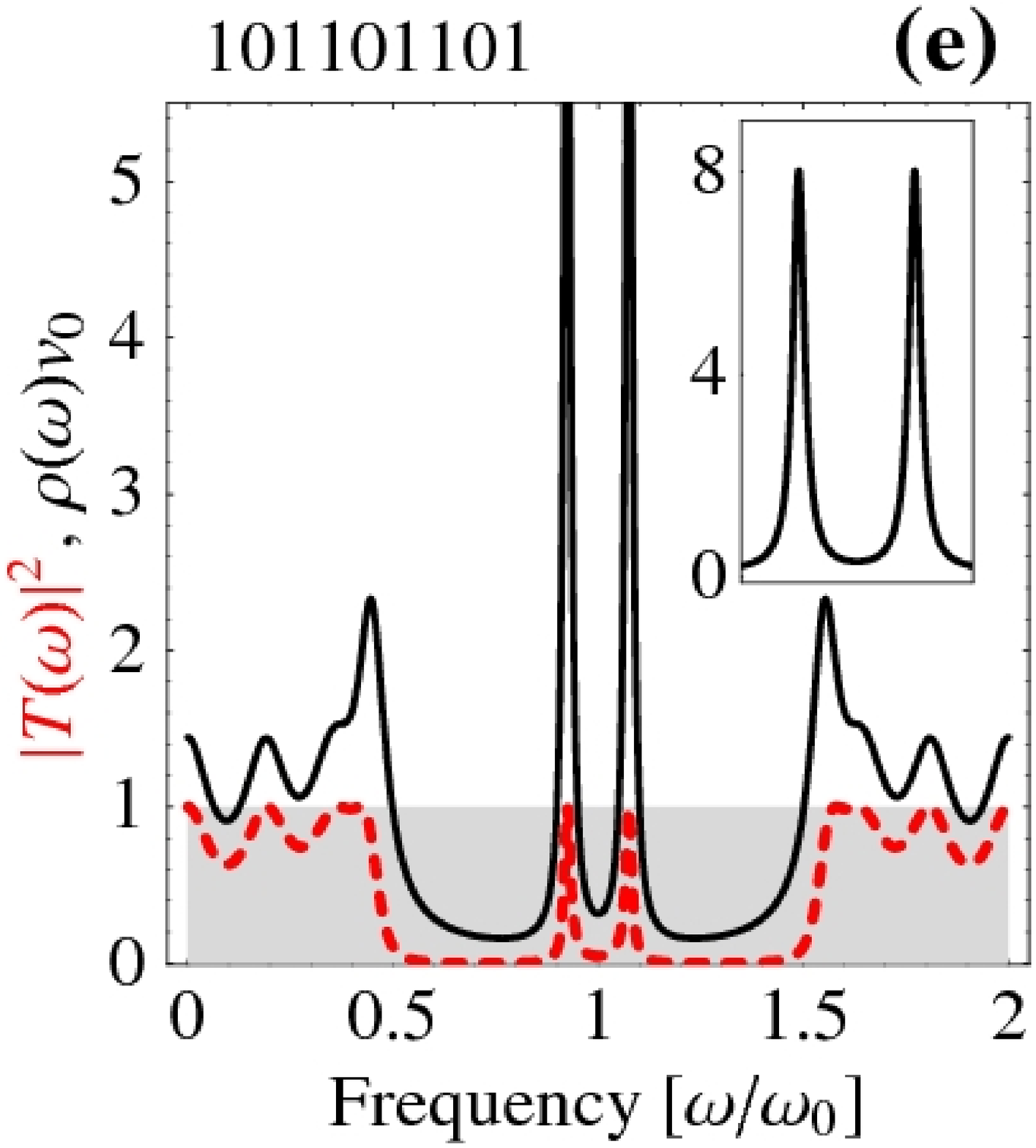}\includegraphics[width=0.33\textwidth]{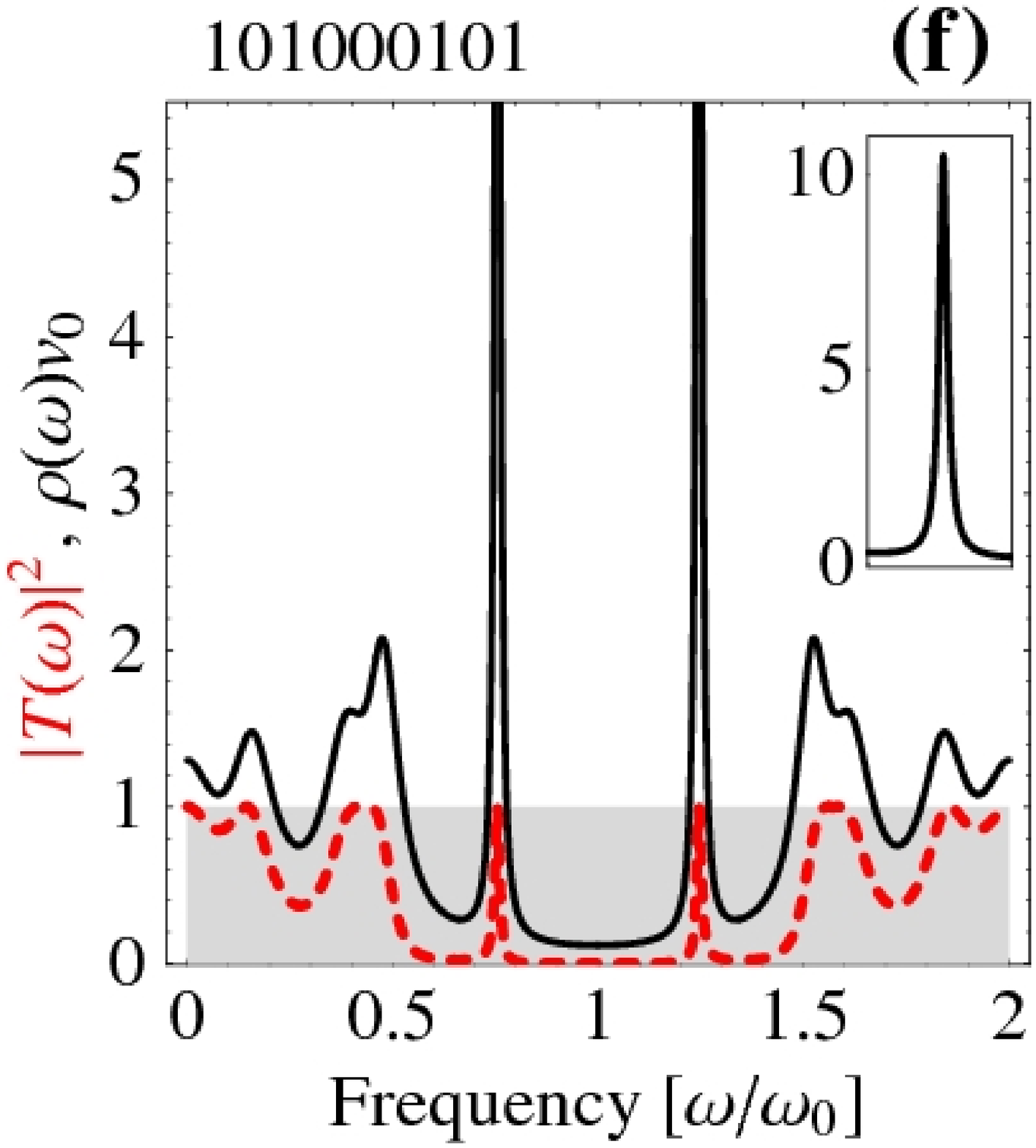}

\caption{(Color online) Transmittance $\left|T(\omega)\right|^{2}$ (dashed
line) and normalized averaged local DOS $\rho(\omega)v_{0}$ {[}as
of Eqs.~\eqref{eq:opt_dos_transmission} and~\eqref{eq:opt_bulkvel},
solid line] for multilayer structures of different geometry: (a) single
$9d_{1}$-thick layer; (b) 3-layer structure; (c) 9-layer periodic
structure; (d) Fabry-P\'erot-like periodic structure with a half-wave
defect; (e) coupled-defect structure; (f) fractal Cantor-like structure.
All multilayers have $N=9$. The vertical scale is chosen alike for
all plots for the ease of comparison, the insets showing the scale
of clipped peaks. The area $[0,1]$ is shaded to show the allowed
region for transmittance, as well as to provide a visual guide for
estimating the integral of~$\rho(\omega)$ {[}see Eq.~\eqref{eq:opt_integral_norm}].
\label{fig:optical_dos}}
\end{figure*}

Physically, the local DOS $\mathcal{N}(\mathbf{r},E)$ is directly
related to the trace of Green's function for the system in question:
$\mathcal{N}(\mathbf{r},E)\propto\textrm{Im Tr }G(\mathbf{r},\mathbf{r},E)$.
By taking the integral Green's function $G(E)$ in place of the local
one, one obtains the value of the DOS~$\mathcal{N}(E)$ that is characteristic
to the whole system for a given value of energy. In a spatially finite
system (a potential well with infinite walls or a closed resonator),
only the states with a discrete set of energy (or frequency) eigenvalues
are allowed. Supposing that these eigenvalues are dense enough, this
integral DOS has a rather loose mathematically but very intuitive
meaning of the number of these discrete states per unit energy. 

In an open resonator, any value of energy corresponds to an eigenstate,
and the DOS transforms into a continuous spectrum~$\mathcal{N}(E)$,
indicative of spectral characteristics for the overall vacuum modification.
It had been shown \cite{Avishai,Witold1,Witold2} that a finite-sized
inhomogeneous potential located in an infinite 1D space causes the
local DOS integrated over the whole space to undergo an overall finite
modification~$\Delta\mathcal{N}(E)\equiv\int_{-\infty}^{\infty}\left[\mathcal{N}(x,E)-\mathcal{N}_{0}(x,E)\right]\mathrm{d}x$.
Note that we are considering a finite-sized inhomogeneity in an infinite
space (an open resonator), as opposed to a finite system isolated
from the outside space (a closed resonator). In the former, $\Delta\mathcal{N}$
was found to be proportional to the derivative of the total phase
accumulated by the wave packet during its transmission through the
inhomogeneity: \begin{equation}
\Delta\mathcal{N}(E)=\left(1/\pi\right)\left(\mathrm{d}\varphi/\mathrm{d}E\right),\label{eq:qm_dos_phase}\end{equation}
which, as seen from Eq.~\eqref{eq:shr_planewave}, becomes asymptotically
zero for very large energies compared to the potential (i.e., if nothing
gets in the particle's way). A similar expression can be used to determine
the local density of electromagnetic states, also called the optical
DOS (for details on its definition in 2D and 3D case, see Refs.~\cite{Busch2D,Giuseppe}).
The transition from local to integral DOS can be made in a similar
manner to the quantum system. In the 1D case (the wave propagation
in a multilayer is a 1D problem when only normal-incidence states
are taken into account) the modification to the optical DOS~$\Delta\mathcal{N}(\omega)$
is also likewise related to the derivative of transmission phase \cite{OptDOSLattices}:\begin{equation}
\Delta\mathcal{N}(\omega)=\left(1/\pi\right)\left(\mathrm{d}(\varphi-\varphi_{0})/\mathrm{d}\omega\right).\label{eq:opt_dos_phase}\end{equation}

The subtraction of the free-space phase shift~$\varphi_{0}$ ensures
that $\Delta\mathcal{N}(\omega)=0$ in free space. In the work by
Barnett and Loudon \cite{BarnettLoudon} it has been shown that the
modification of spontaneous emission rate~$\Gamma$ by inhomogeneous
medium (as compared to the free-space emission rate~$\Gamma_{0}$)
integrated over the whole spectrum must be zero (the Barnett-Loudon
sum rule):\begin{equation}
\int_{0}^{\infty}\left[\frac{\Gamma(\mathbf{r},\omega)}{\Gamma_{0}(\omega)}-1\right]\mathrm{d}\omega=0.\label{eq:opt_BL_liter}\end{equation}

The emission rate~$\Gamma$ is proportional to the local DOS~$\mathcal{N}$,
and the local optical DOS is frequency independent in free space (see
\cite{SVRaman}). Spectral integration in Eq.~\eqref{eq:opt_BL_liter}
can be interchanged with spatial integration over $\Gamma(\mathbf{r},\omega)$.
Hence, a similar relation holds for the integral DOS:\begin{equation}
\int_{0}^{\infty}\Delta\mathcal{N}(\omega)\mathrm{d}\omega=0.\label{eq:opt_BLsumrule}\end{equation}

The transition from local to integral DOS, as well as the transition
from~$\Delta\mathcal{N}$ to~$\mathcal{N}$, involves renormalization
and therefore may be ambiguous. A simple way to counter the divergence
is to accompany the transition from $\varphi-\varphi_{0}$ to~$\varphi$
with normalizing the DOS by the total thickness of the inhomogeneous
medium~$D$ (see \cite{Bendickson}): \begin{equation}
\rho(\omega)\equiv\pi\frac{\Delta\mathcal{N}(\omega)}{D}=\frac{1}{D}\frac{\mathrm{d}\varphi}{\mathrm{d}\omega},\label{eq:opt_bendickson_dos}\end{equation}
The authors in \cite{Bendickson} simply define~$\rho$ as the optical
DOS without going into further details. We can see that it is in fact
neither local nor integral, but rather has the meaning of \emph{local
DOS modified by a finite inhomogeneity, averaged in infinite 1D space}.
In the absence of any inhomogeneity, Eq.~\eqref{eq:opt_bendickson_dos}
gives $\rho(\omega)=1/c$, a known value for the DOS in 1D free space. 

Further, Eq.~\eqref{eq:opt_bendickson_dos} can be used to calculate
$\rho(\omega)$ from the complex transmission coefficient~$T(\omega)$
as\begin{equation}
\rho(\omega)=\frac{\left[\textrm{Im }T(\omega)\right]'\textrm{Re }T(\omega)-\textrm{Im }T(\omega)\left[\textrm{Re }T(\omega)\right]'}{D\left|T(\omega)\right|^{2}},\label{eq:opt_dos_transmission}\end{equation}
the derivation taken with respect to~$\omega$. 

One must realize that the concept of the DOS introduced as in Eqs.~\eqref{eq:qm_dos_phase}--\eqref{eq:opt_dos_phase},
and especially, as in Eq.~\eqref{eq:opt_bendickson_dos}, is not
without controversy. Questions arise already as to the physical meaning
of the quantities involved. For example, one can define a {}``wave
number''~$k$ \emph{a posteriori} from the phase of the transmitted
wave\begin{equation}
k(\omega)=\frac{\varphi}{D}=\frac{\textrm{Arg }T(\omega)}{D},\label{eq:opt_kvector}\end{equation}
which would equal the actual wave number in a homogeneous medium,
or the Bloch wave number in a periodic multilayer at transmission
resonances \cite{Bendickson}. In such special cases, $\rho$~would
equal the inverse group velocity ($\rho(\omega)=\left(\mathrm{d}\omega/\mathrm{d}k\right)^{-1}$),
the latter also equal to the energy velocity. 

In the general case of non-periodic structures, however, the concepts
of phase, group, and energy velocity, as well as their mutual correspondence,
need to be re-examined. For instance, the {}``phase time'' defined
as $\mathrm{d}\varphi/\mathrm{d}\omega$ is, in general, not equal
to the pulse's actual {}``dwell time'' (see \cite{OptDOSLattices,SibiliaDelay}),
although, admittedly, both have a similar frequency dependence and
in some cases the phase time is a very good approximation for the
dwell time \cite{DowlingDelay}. That said, it is safer not to assign
any direct physical meaning to~$k$ defined in Eq.~\eqref{eq:opt_kvector}
in the general case. We will thus treat it like a parameter within
the scope of the present paper.

Another point is that the applicability of 1D models for electromagnetic
problems is in general of limited value. The reason is that reduction
of Maxwell's equations to the scalar wave equation \eqref{eq:helmholtz},
e.g., for multilayered media does not really make the problem entirely
one-dimensional. In reality one deals with finite-width beams rather
than plane waves and with excited atoms that can emit in any direction.
The lateral width of the multilayers is finite, too. As pointed out
in Ref.~\cite{Raymond}, the optical DOS reduces to the expression
\eqref{eq:opt_bendickson_dos} only if off-axis wave propagation is
totally left out.

That kept in mind, the function~$\rho(\omega)$ nevertheless undergoes
the same dramatic variation as does the transmittance itself when
the structure geometry is varied (see Fig.\textbf{~}\ref{fig:optical_dos}).
The peaks in $\left|T(\omega)\right|^{2}$ and $\rho(\omega)$ obviously
correspond to each other. Note that this correspondence is a physical
property of multilayers rather than just a mathematical property of
Eqs.~\eqref{eq:opt_bendickson_dos}--\eqref{eq:opt_dos_transmission}.
Indeed, one can show analytically that $\rho(\omega)\propto\left|T(\omega)\right|^{2}$
for any single layer. This proportionality is due to the phase structure
of the Airy formulas, and is obviously not there for arbitrary $\varphi(\omega)$
in Eq.~\eqref{eq:opt_bendickson_dos}. Numerical calculations confirm
that spectral features in $\left|T(\omega)\right|^{2}$ and $\rho(\omega)$
also correspond for an $N$-layer structure, e.g., as seen in Fig.~\ref{fig:optical_dos}.
It should be possible to show this analytically by induction but it
is outside the scope of the present paper. We note instead that the
same correspondence was observed in higher-dimensional systems (e.g.,
in slab photonic crystals \cite{Ohtaka}).

Besides, one can observe that the sharper is the transmission resonance
around some frequency~$\omega_{r}$, the larger is the value of $\rho(\omega_{r})$.
Sharper transmission resonances correspond to stationary waves with
greater energy localization, and it takes longer for greater energies
to build up inside the structure. Hence, it takes longer for resonant
transmission to manifest in such cases. Therefore the maxima of $\rho(\omega)$
are just those points where delayed light propagation is likely to
be experienced. The DOS spectrum is thus valuable as a quick visual
guide for determining the resonant behavior of any multilayer structure,
as employed earlier \cite{Sibilia}. 

Finally, let us note that although $k$~in Eq.~\eqref{eq:opt_kvector}
cannot be assigned a direct physical meaning in the general case,
it can be used as a parameter, which can provide some heuristic guidance
in experiments on the group velocity dispersion-related effects (e.g.,
pertaining to propagation, compression, delay, and chirp compensation
of ultrashort laser pulses). For some examples involving non-periodic
structures, the reader is referred to Refs.~\cite{Alessandro1,Alessandro2,Liudmila}.

\section{Constraints in multilayers\label{sec:optical}}

In the previous section, the use of~$\rho(\omega)$ defined by Eq.~\eqref{eq:opt_dos_transmission}
as a meaningful characteristic of the structure's optical properties
has been motivated. It was demonstrated that $\rho(\omega)$ can be
strongly modified by altering the geometry of the structure (Fig.~\ref{fig:optical_dos}).
We proceed to show that the degree of geometry-induced modification
imposed on~$\rho$ has fundamental limitations. One of these is the
Barnett-Loudon sum rule -- Eq.~\eqref{eq:opt_BLsumrule} holds both
for the quantum mechanical and for the electromagnetic case when the
corresponding expression for~$\Delta\mathcal{N}(\omega)$ is used.

\begin{figure*}[t]
\hfill{}\includegraphics[width=0.4\textwidth]{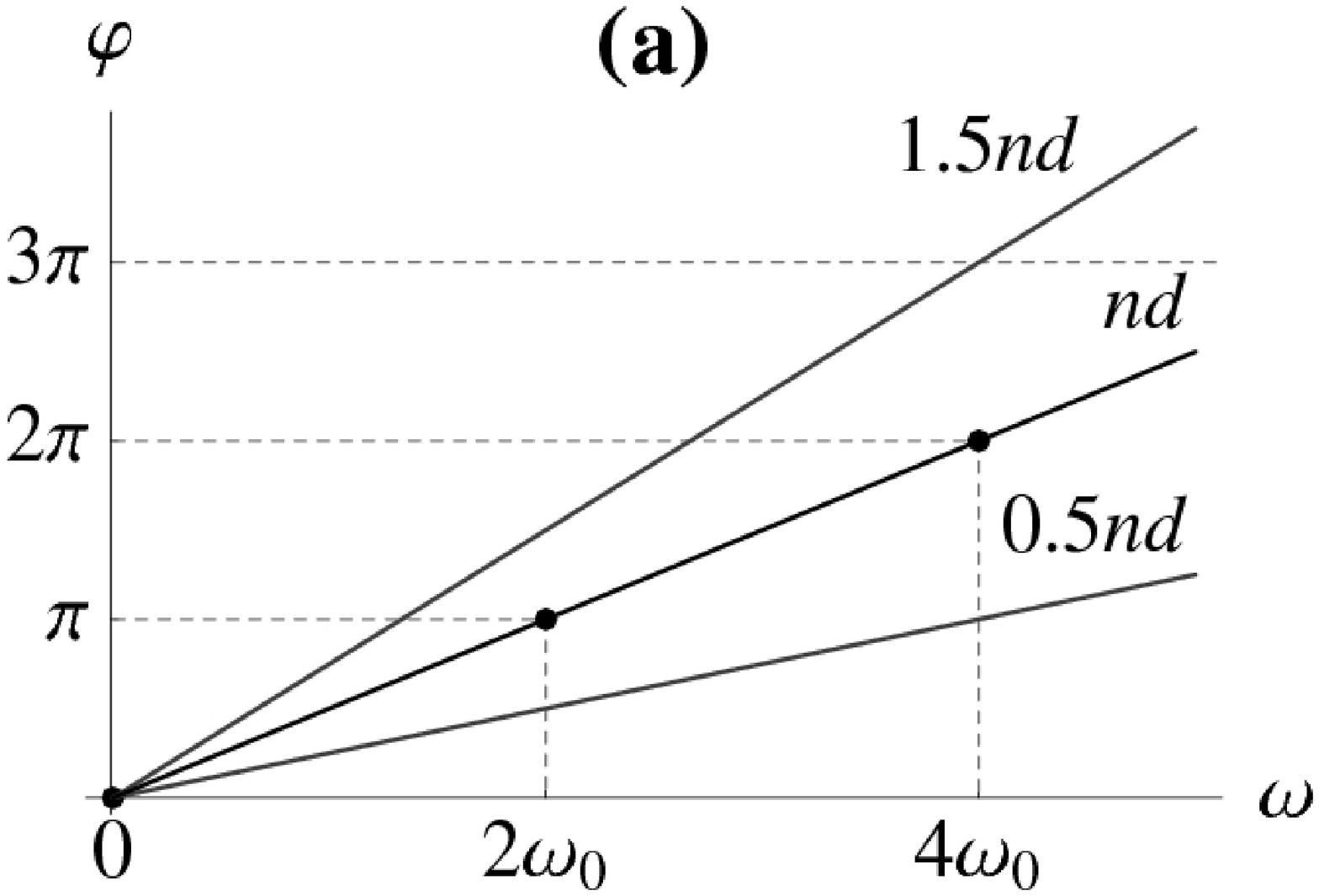}\hfill{}\includegraphics[width=0.4\textwidth]{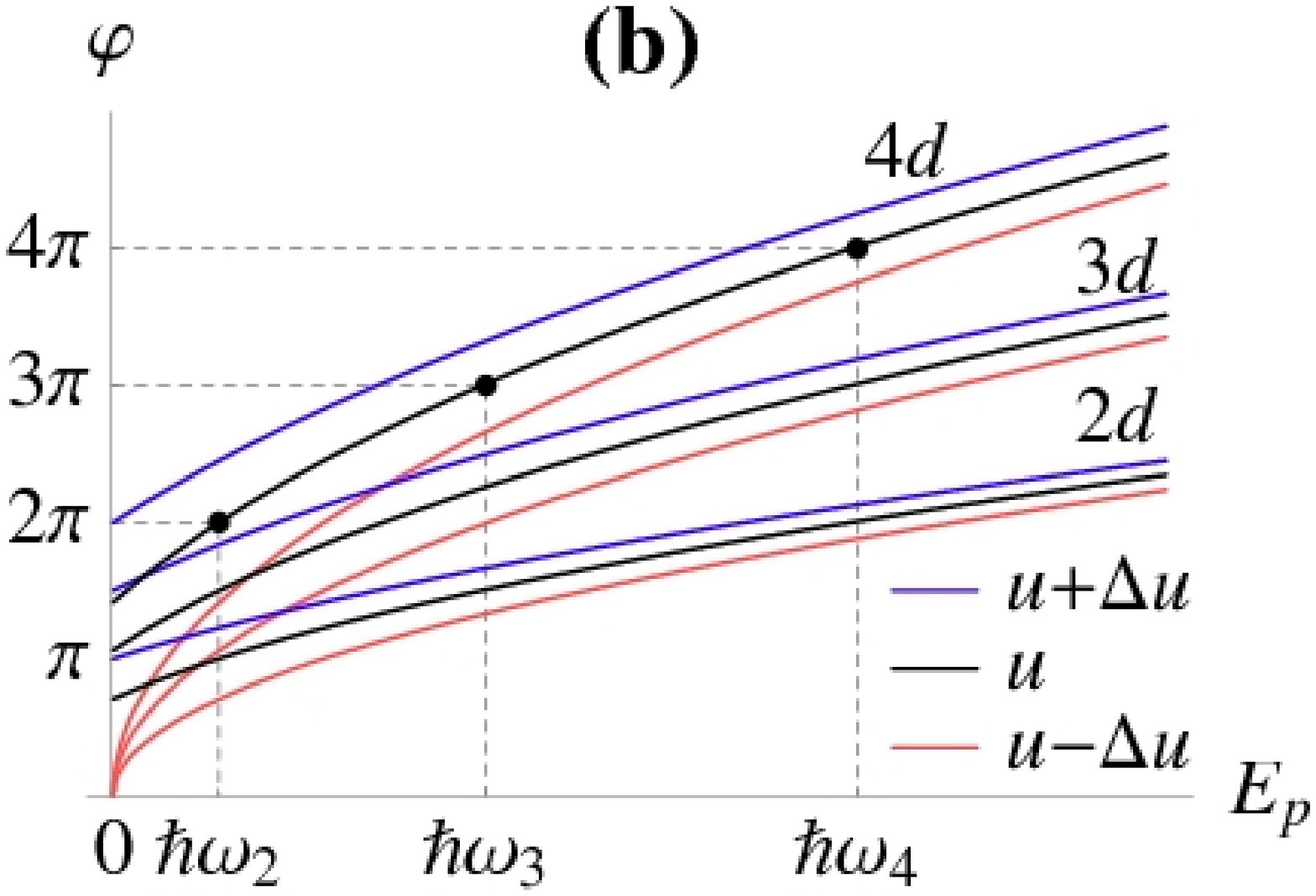}\hfill{}

\caption{(Color online) The schematic frequency dependence of propagation
phase for \textbf{(a)} optical waves {[}Eq.~\eqref{eq:opt_phase}]
and \textbf{(b)} quantum wave function {[}Eq.~\eqref{eq:qm_phase}]
in a slab of homogeneous dielectric and in a length of constant potential,
respectively. \label{fig:dispersion}}
\end{figure*}

If the medium is a QW multilayer, the constraint becomes stricter
and involves integration over \emph{finite} rather than infinite frequency
intervals. As the transmission properties in the SLRF points are given
by Eq.~\eqref{eq:fixed_phase}, the integral of $\rho(\omega)$ between
those points can be evaluated explicitly using Eqs.~\eqref{eq:opt_kvector}
and~\eqref{eq:fixed_phase}--\eqref{eq:fixed_dispersion} (see also
\cite{Nano03}):\begin{equation}
\intop_{2m\omega_{0}}^{2(m+1)\omega_{0}}\rho(\omega)\mathrm{d}\omega=\intop_{k\left[2m\omega_{0}\right]}^{k\left[2(m+1)\omega_{0}\right]}\mathrm{d}k=\frac{N\pi}{D},\label{eq:opt_constraint}\end{equation}
which holds \emph{regardless of the geometrical arrangement} of the
constituent layers in the structure, provided that the layers obey
the relation~\eqref{eq:qwave}. One can further introduce the {}``bulk
velocity prameter'' from the minimum time it takes light to traverse
the multilayer, internal reflections neglected, as \begin{equation}
v_{0}\equiv\frac{D}{\sum_{j}(d_{j}n_{j}/c)}=\frac{2\omega_{0}D}{N\pi}.\label{eq:opt_bulkvel}\end{equation}
This is a parameter independent either of~$\omega$ or of the layer
arrangement of the structure. Making a transition to the dimensionless
frequency $\eta\equiv\omega/\omega_{0}$ and taking into account the
symmetry condition \eqref{eq:qw_mirror}, we arrive at \begin{equation}
\intop_{m}^{m+1}v_{0}\rho(\eta)\mathrm{d}\eta=1\label{eq:opt_integral_norm}\end{equation}
and, further, since $v_{0}=c$ and $\rho_{0}=1/c$ in free space,
\begin{equation}
\intop_{m}^{m+1}\Delta\left[v_{0}\rho(\eta)\right]\mathrm{d}\eta=\intop_{m}^{m+1}\left(v_{0}\rho(\eta)-c\rho_{0}\right)\mathrm{d}\eta=0\label{eq:opt_integral_norm_0}\end{equation}
 for any integer $m\geq0$. 

The conditions \eqref{eq:opt_integral_norm}--\eqref{eq:opt_integral_norm_0}
have no less a universal character than Eq.~\eqref{eq:opt_BLsumrule}.
They physically mean that the modification of the transmission or
dispersion properties due to layer rearrangement in QW multilayers
is only possible within a \emph{finite} frequency range $\omega_{0}$.
One can see in Fig.~\ref{fig:optical_dos} that despite apparently
dramatic modification of~$\rho(\omega)$, the enhancement in one
portion of the spectrum appears compensated by a gap in another portion,
so that the overall DOS, integrated between the SLRF points, remains
unaffected.%
{} It is also important to realize that within~$\omega_{0}$, one can
achieve \emph{any} desired spectral shape, given the sufficient number
of layers and sufficient freedom in their arrangement. For example,
a heuristic optimization algorithm was recently used to demonstrate
that certain aperiodic sequences can be employed to fabricate structures
with desired spectral properties \cite{Subhashish}.

Note, too, the inverse proportionality between $\omega_{0}$~and
the optical path of the constituent layers. It follows that if the
QW condition \eqref{eq:qwave} is broken but the quantities $n_{j}d_{j}$
all remain commensurate, the same reasoning can be applied. Eqs.~\eqref{eq:opt_constraint}
and~\eqref{eq:opt_integral_norm} can then be obtained by subdivision
of the constituent layers, accompanied by the according increase in
the central frequency ($\omega_{0}\to N\omega_{0}$). In the limiting
case of mathematically incommensurate layers, $N$~goes to infinity,
and the structure appears to possess the same freedom as a continuously
inhomogeneous medium would, retaining only asymptotic relation \begin{equation}
\lim_{N\to\infty}\frac{1}{N}\intop_{0}^{N\omega_{0}}\rho(\omega)\mathrm{d}\omega=\frac{\pi}{D},\label{eq:opt_asymptotic}\end{equation}
associated with increasing~$\omega_{0}$ to infinity in Eq.~\eqref{eq:opt_integral_norm_0},
and consequently, representing the sum rule \eqref{eq:opt_BLsumrule}.

\section{Constraints in potential barriers\label{sec:quantum}}

The concept of optical DOS appeared in electrodynamics largely by
the influence of the quantum DOS. Such transfer of concepts makes
use of the analogy between the Helmholtz \eqref{eq:helmholtz} and
Schr\"odinger \eqref{eq:schroedinger} equations, as well as between
a multilayer and a potential barrier, as outlined in Sec.~\ref{sec:grounds}.
In this section, we will attempt to make these analogies work backwards
and determine if, and to what extent, the relation \eqref{eq:opt_constraint}
can be generalized to the quantum mechanical case. 

Consider a binary stepwise potential and the particle with mass~$m_{p}$
and energy $E_{p}\equiv\hbar\omega$ interacting with it in the non-tunneling
regime (Fig.~\ref{fig:structure}b,c). Making use of the expression
\eqref{eq:shr_planewave} for~$k$, we can derive the frequency dependence
for propagation phase of the particle's wave function for constant
potential corresponding to one elementary potential well. Compared
to the same dependence for an optical wave in a homogeneous slab corresponding
to one constituent layer, it has the form ($\alpha\equiv\sqrt{2m_{p}}/\hbar$)
\begin{eqnarray}
\varphi^{\textrm{(qm)}}(\omega) & \equiv & k^{(\textrm{qm})}(\omega)d=d\alpha\sqrt{\hbar\omega+u};\label{eq:qm_phase}\\
\varphi^{\textrm{(opt)}}(\omega) & = & nd\omega/c.\label{eq:opt_phase}\end{eqnarray}

Fig.~\ref{fig:dispersion} schematically shows both these dependencies.
In the optical case (Fig.~\ref{fig:dispersion}a) the only variable
parameter is the slope given by $nd$. Hence, with the aid of Eq.~\eqref{eq:qwave}
it becomes possible to achieve exactly the same dispersion relation,
and hence the same set of SLRF points, for both constituent layers
when $n_{0}\ne n_{1}$. This is what forms the foundation for reasoning
presented in the previous section. In the quantum case (Fig.~\ref{fig:dispersion}b)
$u$~and~$d$ are seen to contribute in an essentially different
way. Is is thus not possible to arrive at the same dispersion relation
for two different potentials ($u_{0}\neq u_{1}$).

\begin{table*}[!t]

\caption{Some values of~$\beta$ that allow integer solutions of Eq.~\eqref{eq:qm_quadratic},
along with some of such solutions obtained numerically. \label{tab:numbers}}

\noindent \begin{tabular}{cc|cc}
\hline 
$\beta<1$&
$\left(j_{\textrm{qw}}:j_{0},j_{1}\right)$&
$\beta\ge1$&
$\left(j_{\textrm{qw}}:j_{0},j_{1}\right)$\tabularnewline
\hline
1/4&
$(1:13,15)$; $(1:181,209)$; $(2:122,126)$&
1&
$(2:7,8)$; $(2:26,30)$; $(3:17,18)$; $(3:99,105)$; $(4:31,32)$\tabularnewline
1/3&
$(1:9,11)$; $(1:89,109)$; $(2:90,94)$&
3/2&
$(2:8,10)$; $(2:68,86)$; $(3:63,69)$\tabularnewline
1/2&
$(1:5,7)$; $(1:29,41)$; $(2:58,62)$; $(3:207,213)$&
2&
$(2:10,14)$; $(2:58,82)$; $(3:45,51)$\tabularnewline
2/5&
$(1:7,9)$; $(1:55,71)$; $(2:74,78)$&
4&
$(3:7,9)$; $(3:18,24)$; $(4:14,16)$; $(4:52,60)$; $(5:23,25)$\tabularnewline
2/3&
$(1:3,5)$; $(1:11,19)$; $(2:42,46)$; $(3:153,159)$&
6&
$(3:9,15)$; $(3:33,57)$; $(4:16,20)$\tabularnewline
4/5&
$(1:2,4)$; $(1:5,11)$; $(1:13,19)$; $(2:34,38)$&
10&
$(4:8,12)$; $(4:32,52)$; $(5:35,45)$\tabularnewline
\hline
\end{tabular}
\end{table*}
\begin{figure*}[!t]
\includegraphics[width=0.33\textwidth]{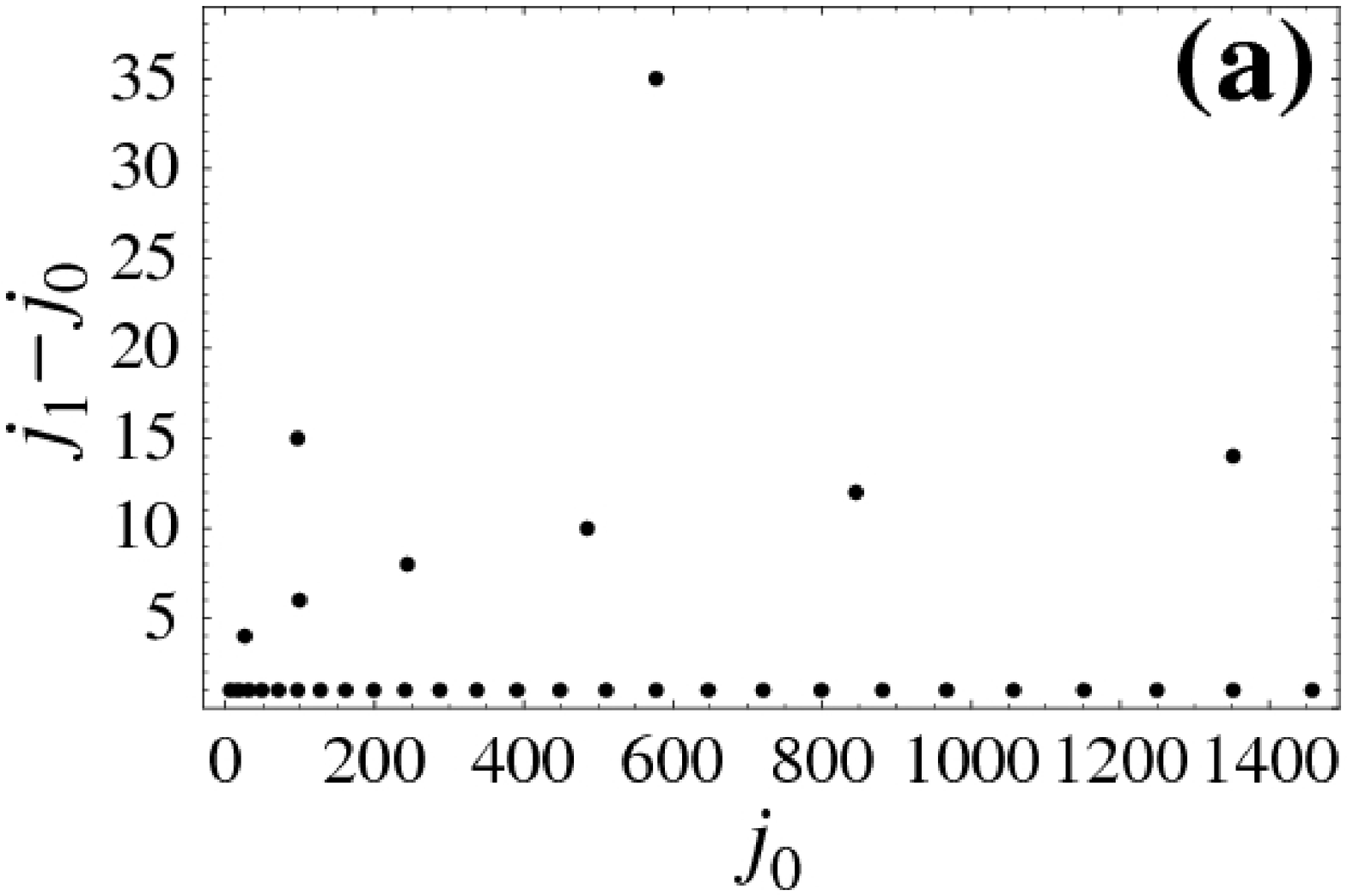}\includegraphics[width=0.33\textwidth]{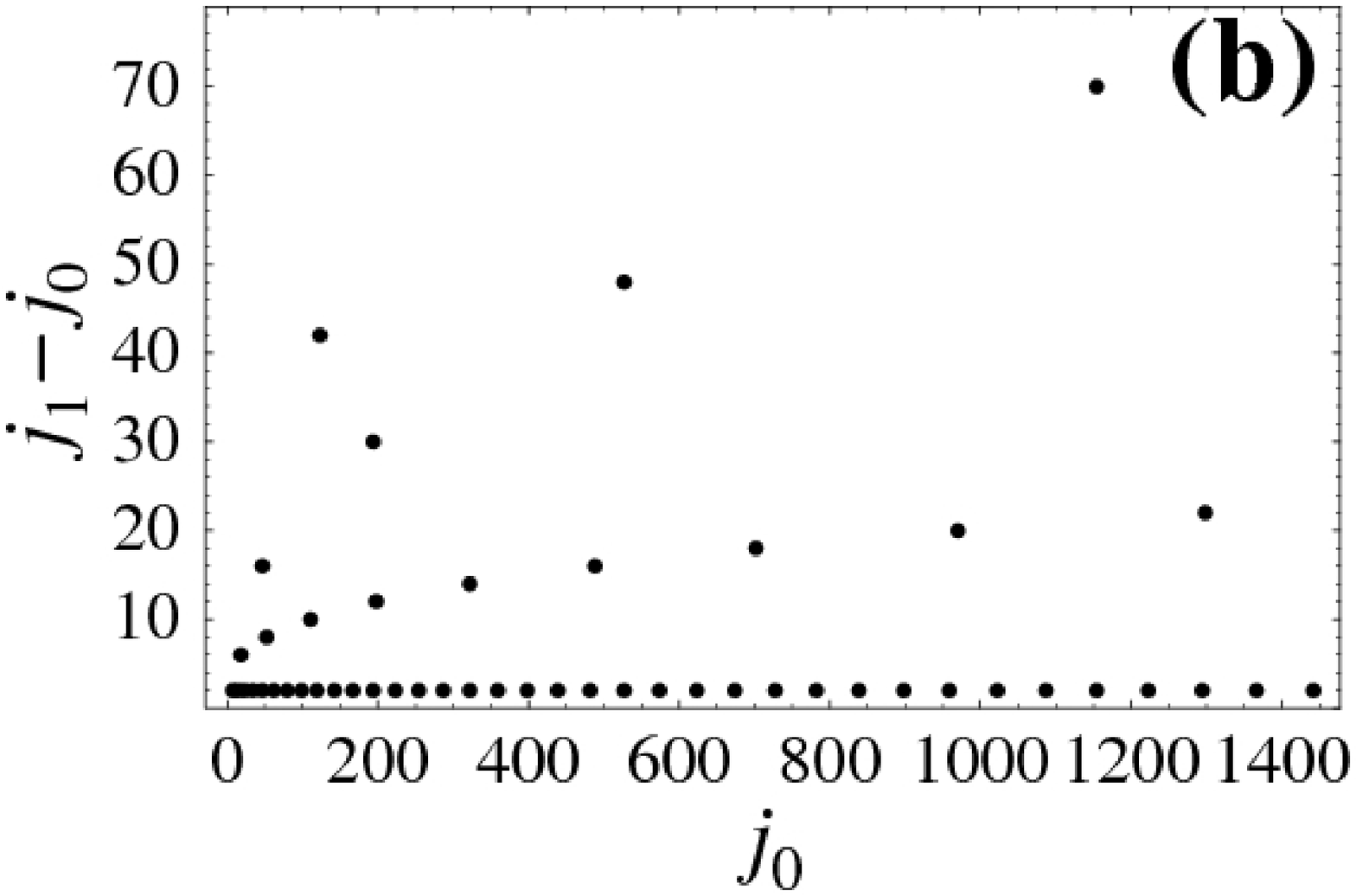}\includegraphics[width=0.33\textwidth]{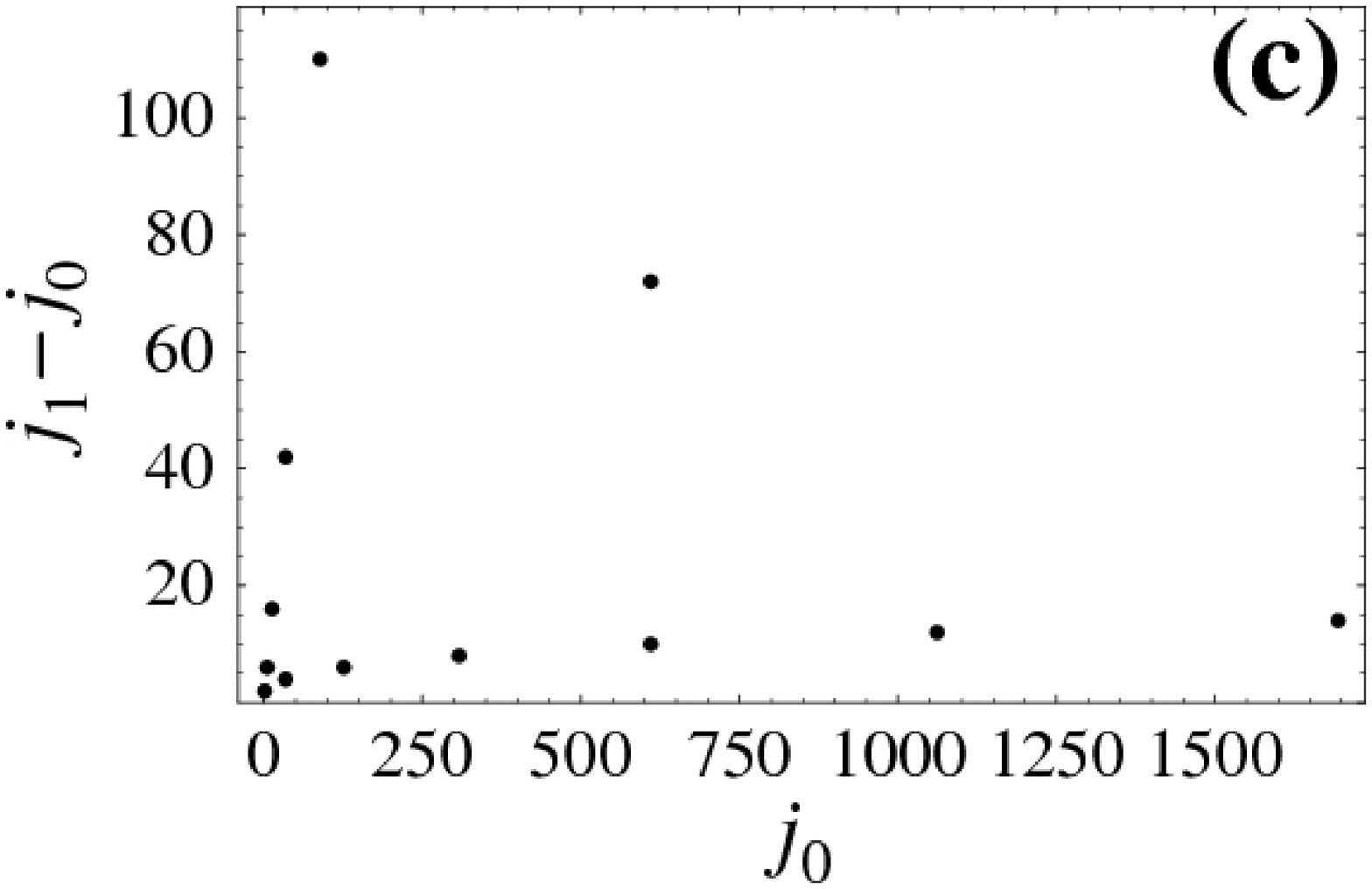}

\caption{The distribution of integer solutions~$j_{0,1}$ of Eq.~\eqref{eq:qm_quadratic}
for \textbf{(a)} $\beta=1$, \textbf{(b)} $\beta=4$, and \textbf{(c)}
$\beta=4/5$. Distinct solution groups ({}``branches'') can be seen.\label{fig:numbers}}
\end{figure*}

However, one can still define a set of frequency points (though no
longer equidistant) where $\varphi^{(\textrm{qm})}(\omega_{j})=j\pi$.
In these points, as can be seen from Eqs.~\eqref{eq:shr_planewave}
and~\eqref{eq:qm_barrier}, the whole structure would be totally
{}``transparent'' for incoming quantum particles (the Ramsauer effect).
If the structure is binary, the frequencies for both kinds of elementary
wells are given by \begin{equation}
\hbar\omega_{j_{0,1}}^{(0,1)}=j_{0,1}^{2}\pi^{2}/d_{0,1}^{2}\alpha^{2}-u_{0,1}.\label{eq:qm_SLRF}\end{equation}

Since two different parabolic curves can still have intersection points,
one can manage to achieve $\omega_{j_{0}}^{(0)}=\omega_{j_{1}}^{(1)}$
for two pairs of $j_{0}$~and~$j_{1}$. The reasoning presented
in the previous section can then be reproduced involving the quantity~$\rho$
defined exactly as in Eqs.~\eqref{eq:opt_bendickson_dos} and~\eqref{eq:opt_constraint}
and having the same physical background. The dependence on~$\omega$,
however, will be more complex due to inherent dispersion as seen in
Eq.~\eqref{eq:shr_planewave}. 

For simplicity and for the sake of further analogy between optical
and quantum systems, let us require one of the equal frequency pairs
in Eq.~\eqref{eq:qm_SLRF} to correspond to $\omega=0$ (which is
always true for optical waves where all dispersion curves pass trough
the origin, see Fig.~\ref{fig:dispersion}a). In this case we arrive
at \begin{equation}
\alpha^{2}d_{0}^{2}u_{0}=\alpha^{2}d_{1}^{2}u_{1}=j_{\textrm{qw}}^{2}\pi^{2},\label{eq:qm_static}\end{equation}
which can be seen as a quantum analogy to the condition~\eqref{eq:qwave}.
The second pair $\left(j_{0},j_{1}\right)$ can then be found as an
integer solution of the equation (first suggested in our earlier work
\cite{Nano07})\begin{equation}
\begin{array}{c}
j_{0}^{2}-\left(1-\beta/j_{\textrm{qw}}^{2}\right)j_{1}^{2}=\beta,\\
j_{0,1}>j_{\textrm{qw}};\quad\beta\equiv(u_{0}-u_{1})\alpha^{2}d_{0}^{2}/\pi^{2}.\end{array}\label{eq:qm_quadratic}\end{equation}

It can be seen that for any integer $j_{0,1}>j_{\textrm{qw}}$ there
is a rational~$\beta$ that solves Eq.~\eqref{eq:qm_quadratic}.
But $\beta$~is related to the parameters of the constituent potential
wells. So, the inverse problem, i.e, finding suitable~$j_{0,1}$
for a given~$\beta$, is far more interesting from a physical point
of view. However, is not so straightforward and is mathematically
related to finding Pythagorean triples in integer numbers. One can
confirm numerically that there are a multitude of solutions for many
rational values of~$\beta$ (see Table~\ref{tab:numbers}). Some
of them can be represented via recurrent relations, e.g., for $\beta=1$
some of the solutions represent a series: \begin{equation}
j_{1}^{(i+1)}=j_{1}^{(i)}+6+4i,\quad j_{0}^{(i)}=j_{1}^{(i)}-1,\label{eq:qm_one_series}\end{equation}
where $j_{1}^{(0)}=j_{\textrm{qw}}^{(\textrm{min})}=2$. Other cases
are more complex, but they, too, can be seen to form distinct solution
branches (Fig.~\ref{fig:numbers}).

Once $j_{0}$~and~$j_{1}$ have been found, an analogous relation
to Eq.~\eqref{eq:opt_constraint} can be formulated as \begin{equation}
\intop_{{0=\hbar\omega}_{j_{\textrm{qw}}}}^{\hbar\omega_{j_{0}}^{(0)}=\hbar\omega_{j_{1}}^{(1)}}\rho(E)\mathrm{d}E=\frac{\hbar\pi}{D}\left[N_{0}(j_{0}-j_{\textrm{qw}})+N_{1}(j_{1}-j_{\textrm{qw}})\right].\label{eq:qm_constraint}\end{equation}

Note that Eq.~\eqref{eq:qm_constraint} is more complicated than
its optical counterpart \eqref{eq:opt_constraint}, and becomes, in
general, dependent on the number of constituent potential wells $N_{1,2}$.
This dependence cannot be eliminated because one sees from Eq.~\eqref{eq:qm_quadratic}
that it is impossible to have $j_{0}=j_{1}$ without violating the
assertion that $j_{0,1}>j_{\textrm{qw}}$. It is still, however, completely
independent of layer rearrangement. In this sense, Eq.~\eqref{eq:qm_quadratic}
represents a universal quantum mechanical conservation relation for
the DOS over a finite energy interval.

To demonstrate the results numerically, we have considered a stepwise
AlAs/GaAs quantum well ($\Delta u=1000$~meV). To aim at $\beta=4/5$,
we have taken $d_{0}=21.2$~\AA, $d_{1}=47.4$~\AA, according
to Eqs.~\eqref{eq:qm_static}--\eqref{eq:qm_quadratic}. One possible
solution of Eq.~\eqref{eq:qm_quadratic} would then be $j_{\textrm{qw}}=1$,
$j_{0}=2$, $j_{1}=4$ (see Table~\ref{tab:numbers}). The structures
made of $N=9$ elementary wells were used, and the numbers $N_{0}$~and~$N_{1}$
were fixed, too, at the values 4 and 5, respectively.

\begin{figure*}
\includegraphics[
width=0.33\textwidth]{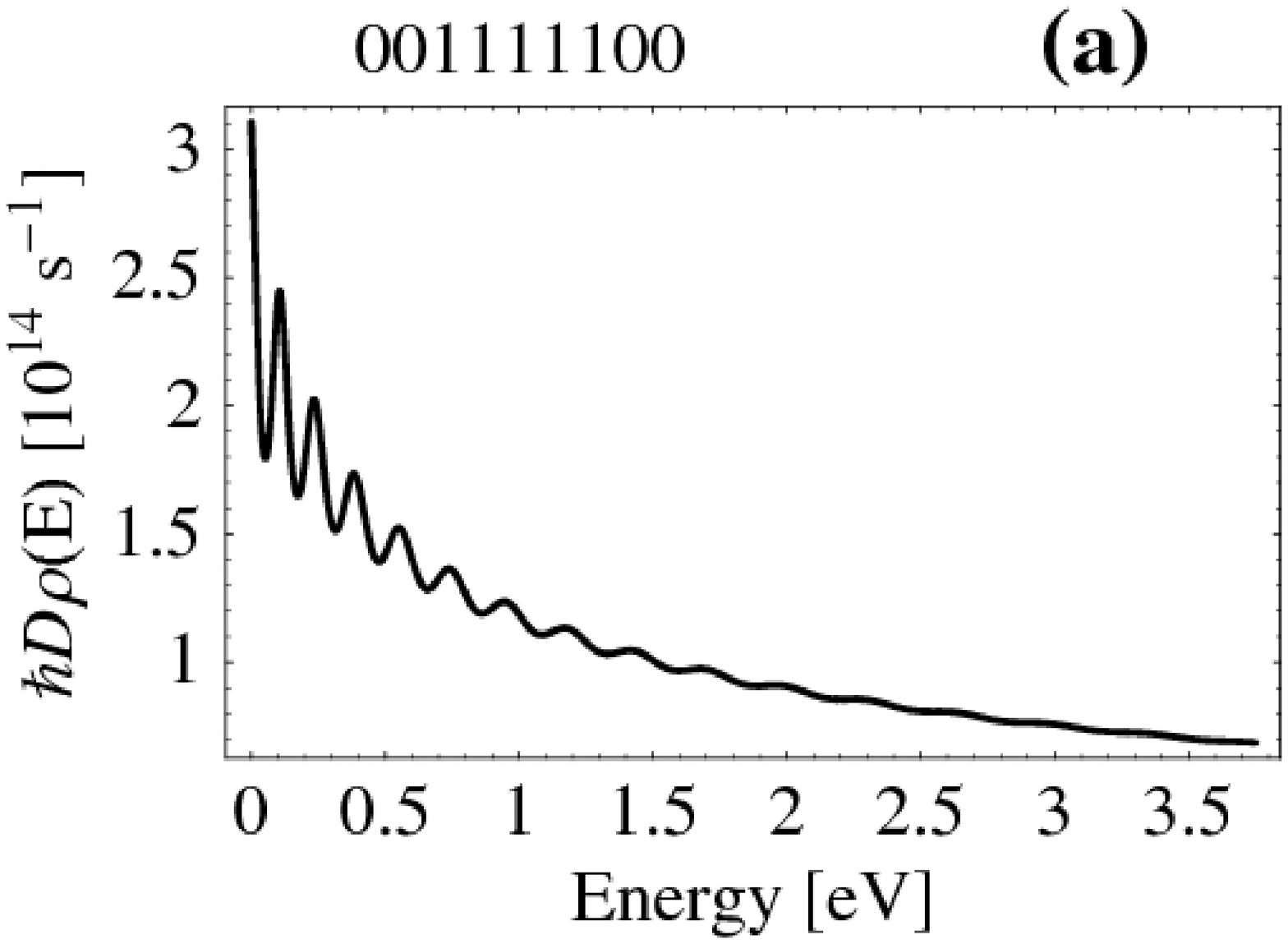}
\includegraphics[
width=0.33\textwidth]{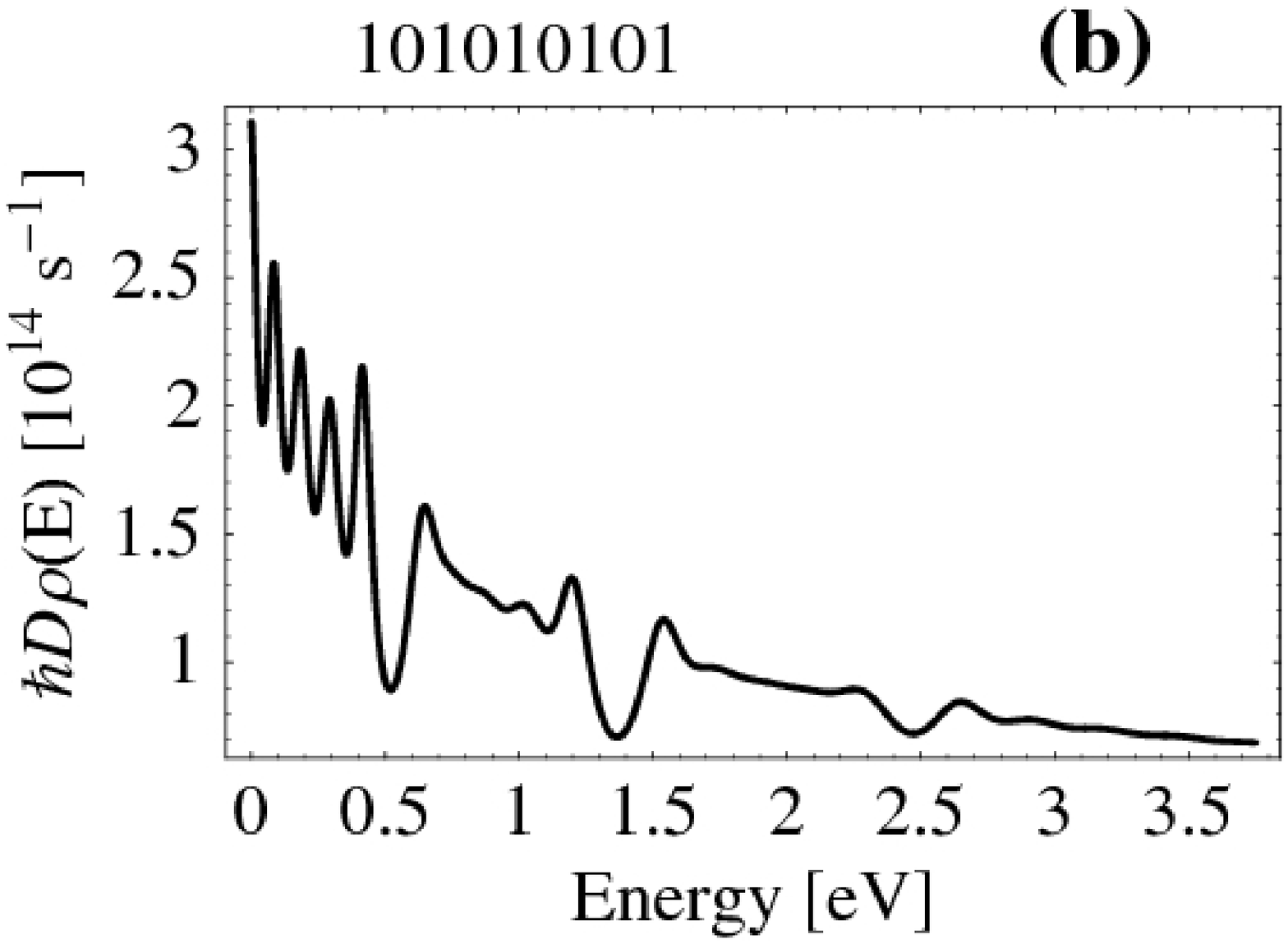}
\includegraphics[
width=0.33\textwidth]{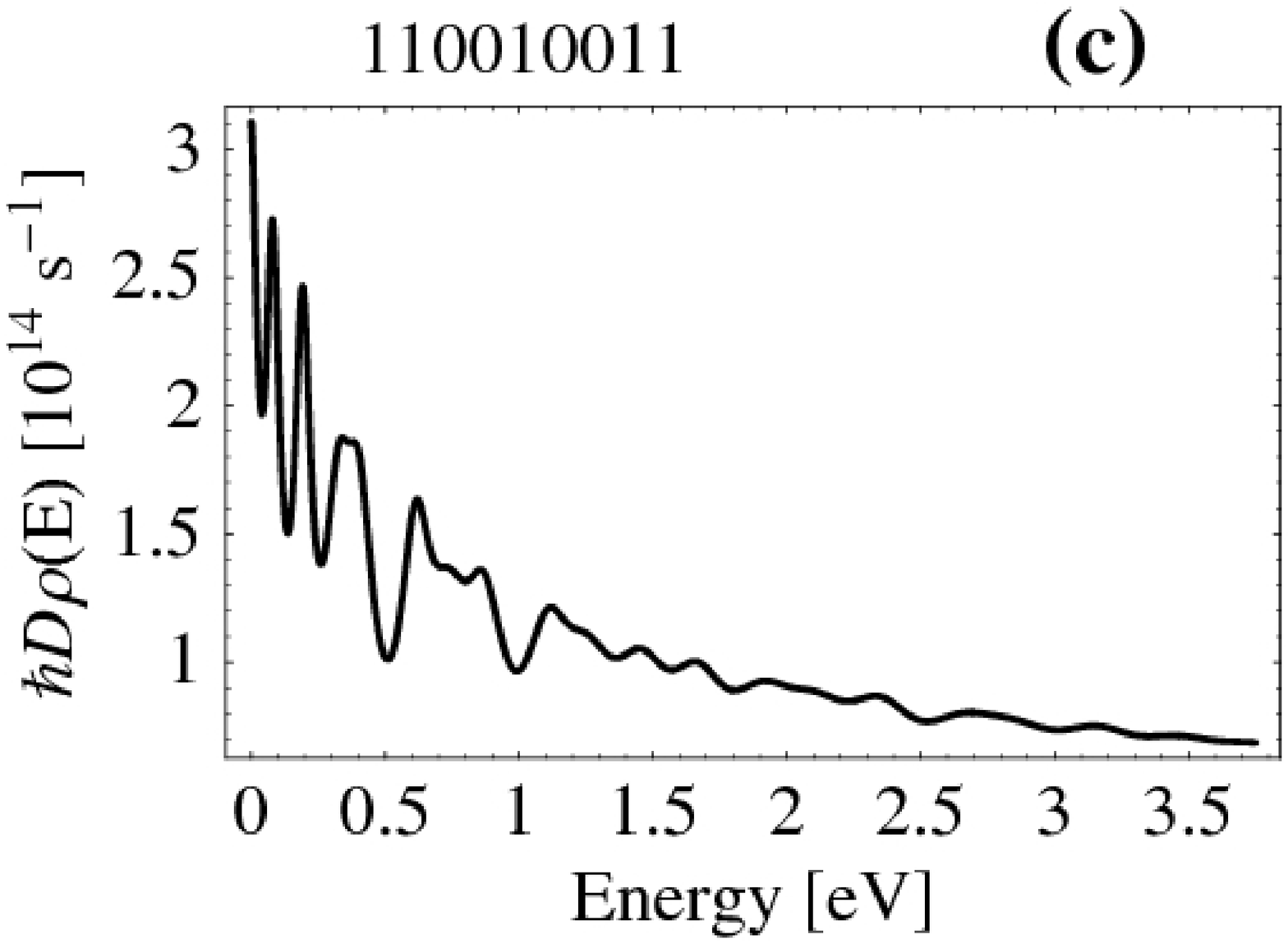}

\includegraphics[
width=0.33\textwidth]{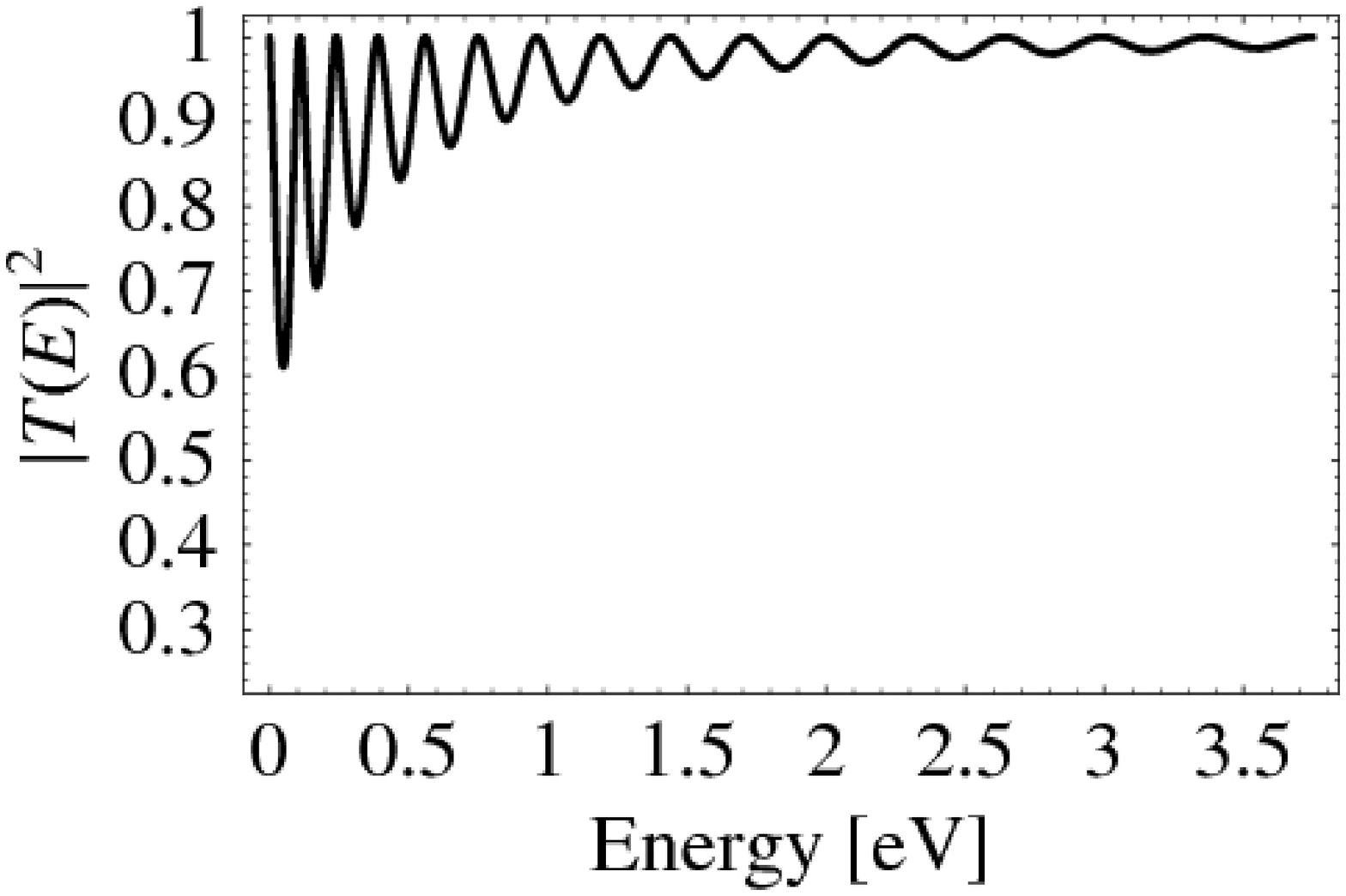}
\includegraphics[
width=0.33\textwidth]{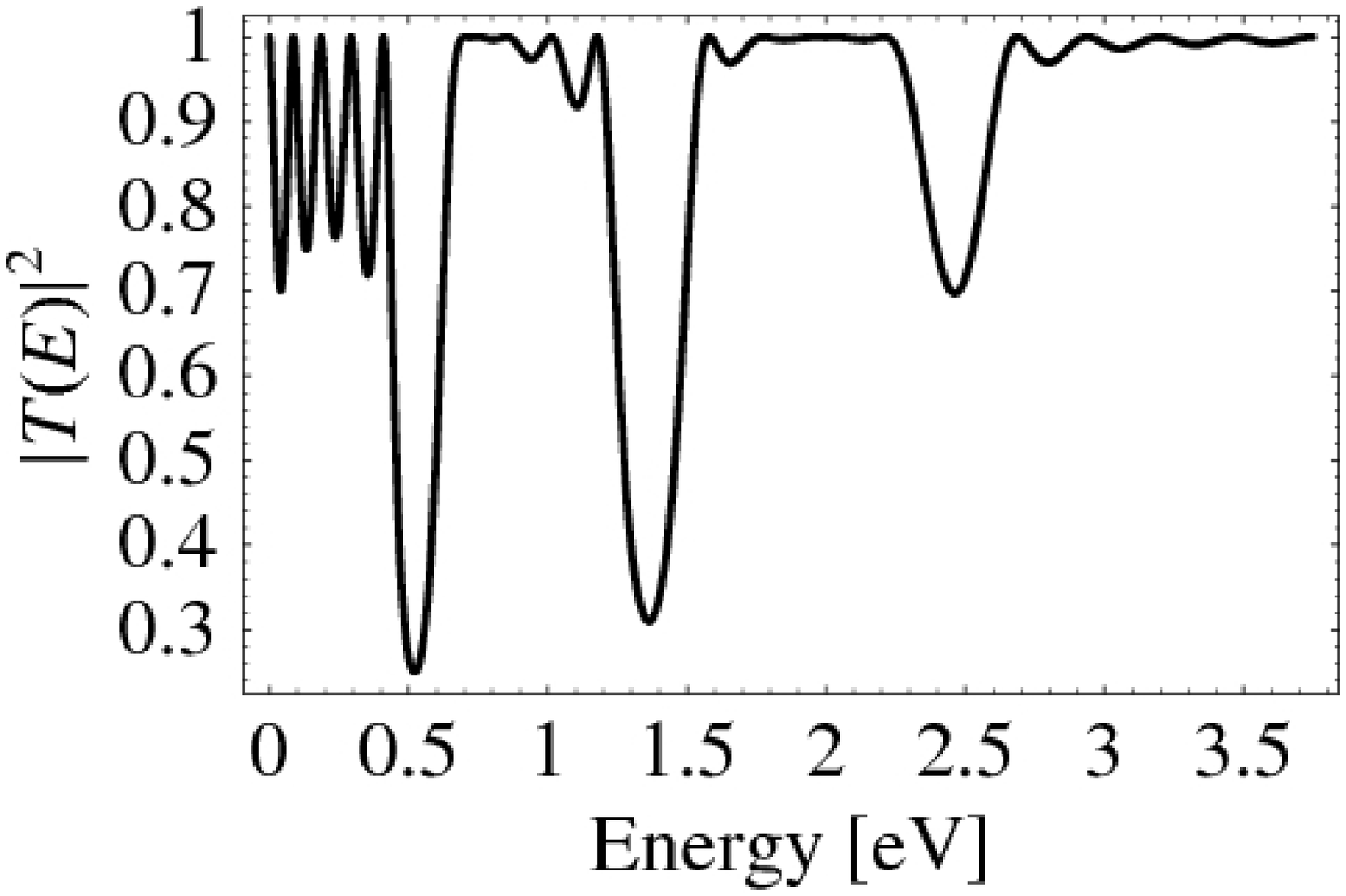}
\includegraphics[
width=0.33\textwidth]{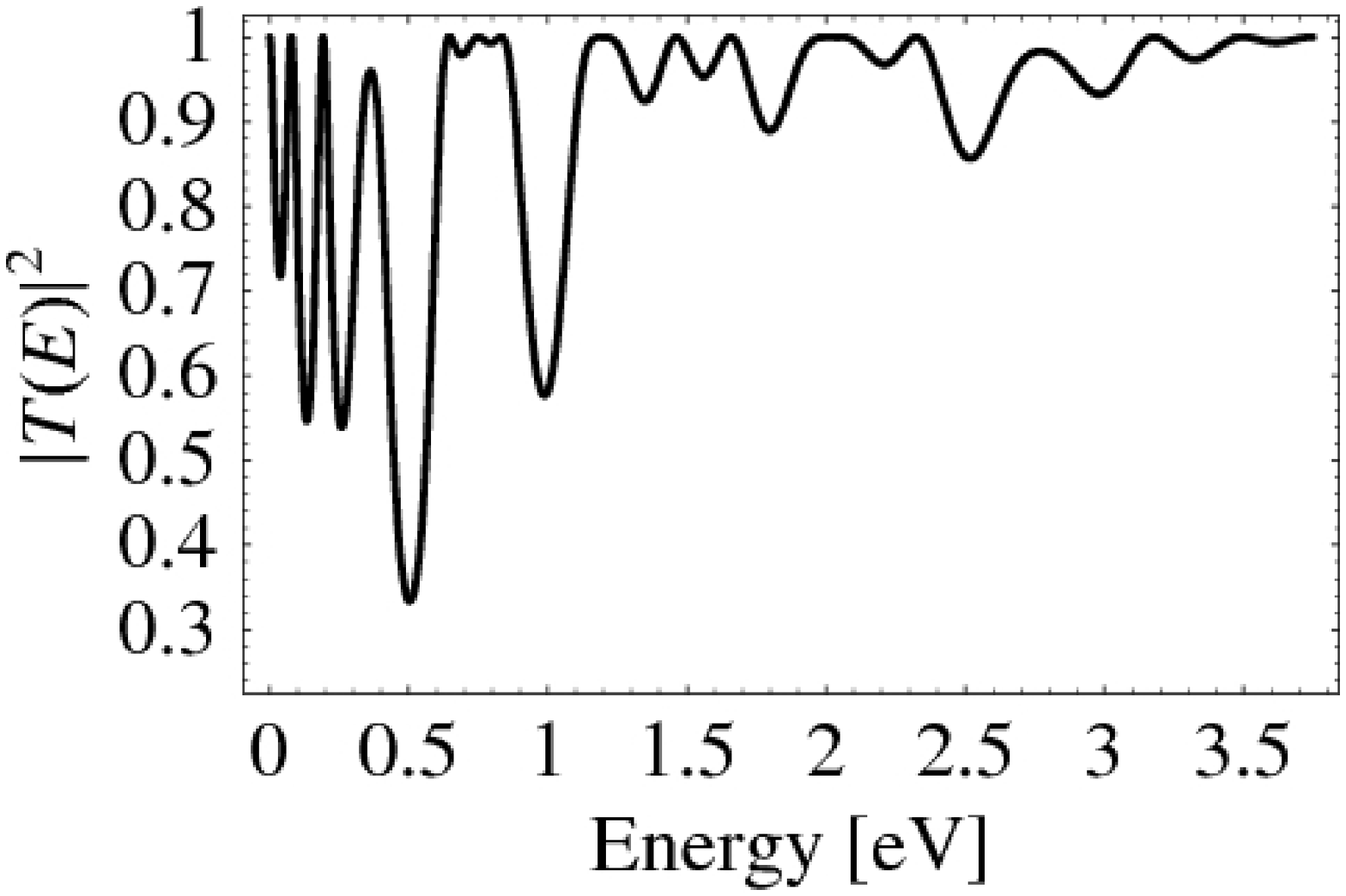}

\caption{The quantum averaged DOS $\rho(E)$ (top) and transmittance $\left|T(E)^{2}\right|$
(bottom) for an AlAs/GaAs quantum well and $\beta=4/5$ in three stepwise
potential wells differing only by the elementary well rearrangement:
\textbf{(a)} single-layer structure; \textbf{(b)} periodic structure;
\textbf{(c)} non-periodic structure. The portion between the SLRF
points {[}0\ldots{}3.75~eV] is depicted, which corresponds to a
solution of Eq.~\eqref{eq:qm_quadratic} for $j_{\textrm{qw}}=1$,
$j_{0}=2$, $j_{1}=4$.\label{fig:quantum_dos}}
\end{figure*}

We see in Fig.~\ref{fig:quantum_dos} that both~$\rho(E)$ and the
transmission~$T(E)$ are subject to quite a strong modification.
It resembles the modification seen in dielectric multilayers (compare,
e.g., Fig.~\ref{fig:quantum_dos}a,b with Fig.~\ref{fig:optical_dos}a,c).
Two differences are the presence of a decaying background due to the
inherent dispersion {[}see Eq.~\eqref{eq:qm_phase}] and the lack
of periodicity because Eq.~\eqref{eq:qwave} can no longer be satisfied. 

However, if we integrate $\rho(E)$ between the SLRF points ($\hbar\omega_{j_{\textrm{qw}}}=0$
and $\hbar\omega_{j_{0}}=\hbar\omega_{j_{1}}=3.75\:\textrm{eV}$)
as provided by Eq.~\eqref{eq:qm_quadratic}, we can see that the
integral does not change when the layers are rearranged. Table~\ref{tab:integral_dos}
provides the results for nine different structures and for several
upper integration limits. It can be seen that both below and above
3.75~eV the integrals vary from structure to structure. When, however,
the correct integration limits are chosen, the difference vanishes
and all integrals equal 19, which is the right-hand side of Eq.~\eqref{eq:qm_constraint}
for the chosen values of parameters. %
{}

\begin{table}[b]

\caption{Numerically evaluated integrals $(D/\hbar\pi)\int\rho(E)\mathrm{d}E$
{[}as in Eq.~\eqref{eq:qm_constraint}] from 0 to several upper energy
values for nine structures with $N=9$, $N_{0}=4$, and $N_{1}=5$
(same as in Fig.~\ref{fig:quantum_dos}). Standard deviation of the
values across all structures for each upper integration limit is provided
in the lowest row. The limit of 3.75~eV (the obtained value of the
SLRF point) is accompanied by a drop in standard deviation down to~$10^{-8}$,
which falls within accuracy limits for numerical integration. \label{tab:integral_dos}\medskip{}
}

\noindent \begin{tabular}{c|cccc}
\hline 
Structure&
0\ldots{}1 eV&
3 eV&
3.75 eV&
\ldots{}5 eV\tabularnewline
\hline
001111100&
7.5531&
16.4036&
19.0000&
22.8568\tabularnewline
010111100&
7.5526&
16.3991&
19.0000&
22.8593\tabularnewline
100111100&
7.5481&
16.3982&
19.0000&
22.8582\tabularnewline
110011100&
7.5207&
16.4016&
19.0000&
22.8574\tabularnewline
010111010&
7.5890&
16.3991&
19.0000&
22.8600\tabularnewline
100110011&
7.5177&
16.4048&
19.0000&
22.8592\tabularnewline
100111001&
7.5198&
16.4017&
19.0000&
22.8512\tabularnewline
110010011&
7.5373&
16.3996&
19.0000&
22.8603\tabularnewline
101010101&
7.5880&
16.3982&
19.0000&
22.8657\tabularnewline
\hline
Std. deviation&
0.027&
0.0045&
$8\times10^{-9}$&
0.0038\tabularnewline
\hline
\end{tabular}
\end{table}

\section{Discussion\label{sec:discussion}}

The equations \eqref{eq:opt_constraint}--\eqref{eq:opt_integral_norm_0}
and~\eqref{eq:qm_constraint} constitute the main result of this
paper pertaining to optical and electronic heterostructures, respectively.
In both cases, we are dealing with conservation of the DOS~$\rho$
integrated across a finite energy or frequency region. As discussed
in Sec.~\ref{sec:dos}, $\rho$~represents the averaged local DOS
as modified by the presence of finite-sized inhomogeneous structure
in an infinite 1D free space. It is related to the dispersion and
the transmission properties of the heterostructures in question {[}see
Eqs.~\eqref{eq:opt_bendickson_dos}--\eqref{eq:opt_dos_transmission}]. 

These properties, as well as the DOS, can undergo dramatic modification
as compared to those of homogeneous media (see Figs.~\ref{fig:optical_dos}
and~\ref{fig:quantum_dos}) because a multilayer structure or a stepwise
potential barrier can be very complex. Nevertheless, the modification
appears to be limited both in its amount {[}see the right hand side
of Eqs.~\eqref{eq:opt_constraint} and~\eqref{eq:qm_constraint}]
and in its extent (by the finite integration limits in those equations). 

There is an elegant physical explanation for the relations obtained.
By engineering the geometrical properties of an inhomogeneous structure,
it is only posible to redistribute the available electromagnetic or
quantum states across the spectrum, but impossible to alter the {}``total
number'' of the states. The latter turns out to be related to the
size or {}``1D volume'' of the structure {[}as seen by the presence
of~$N$ at the right-hand side of Eqs.~\eqref{eq:opt_constraint}
and~\eqref{eq:qm_constraint}] and represents integrated characteristics
of the structure-affected vacuum. So, an enhancement of the DOS in
some pars of the spectrum (like the band edge resonances for a periodic
structure in Fig.~\ref{fig:optical_dos}c) giving rise, e.g., to
the spontaneous emission enhancement, is inevitably accompanied by
a suppression of the DOS in other spectral region (like the band gap
in the same figure), leading to the inhibition of light propagation
and all phenomena involving light-matter interaction \cite{PCBooks}.

In this sense, the results obtained resemble already known constraints
on the DOS like the Barnett-Loudon sum rule \eqref{eq:opt_BLsumrule}.
However, in the relations obtained in this work the integration involved
is \emph{finite} rather than infinite. For the optical case, this
means a tighter restriction on the spectral redistribution of the
DOS. The compensation of suppression snd enhancement must occur within
the frequency interval~$\omega_{0}$. This interval is determined
solely by the elementary constituent element of the structure in question
{[}see Eq.~\eqref{eq:qwave}]. It is totally independent of geometrical
arrangement of these elements. In other words, the QW condition \eqref{eq:qwave}
enforces the existence of certain points in the spectrum (the SLRF
points) that cannot be {}``transcended'' by electromagnetic states
that are {}``pushed around'' the spectrum by layer rearrangement.

On the other hand, the spectral properties of the structure can be
arbitrary everywhere between the SLRF points~\eqref{eq:fixed_phase}.
It should also be noted that the increase of~$N$ causes the details
in the spectra to become finer, and the variation of $T(\omega)$~and~$\rho(\omega)$
to get more rapid. These results can help to understand the underlying
physics of complex media.

Similar conclusions can be formulated for a quantum particle in a
stepwise potential barrier. However, the inherent quadratic dispersion
as seen in Eqs.~\eqref{eq:shr_planewave} and~\eqref{eq:qm_phase}
results in many differences. First and foremost, the SLRF points are
no longer guaranteed. Not only a relative restriction on constituent
elements \eqref{eq:qm_static} analogous to the QW condition \eqref{eq:qwave}
is required, but also individual constraints on $u_{0,1}$ and $d_{0,1}$
are necessary, so as to provide special values of~$\beta$ as determined
by Eq.~\eqref{eq:qm_quadratic}. This makes the binarity of the structure
an important requirement in contrast to the optical case where Eqs.~\eqref{eq:qwave}
and~\eqref{eq:opt_phase} can be extended to as many kinds of constituent
elements as needed. Because the equation~\eqref{eq:qm_quadratic}
is quadratic rather than linear, the SLRF points occur far more seldom
than in the optical case and are no longer equidistant. However, they
still do occur on a regular basis if they occur at all for a given
choice of parameters, as seen in Fig,~\ref{fig:numbers}. In this
way, we have provided a way for engineering an electronic heterostructure
where the DOS modification due to structure complexity is confined
in a finite spectral region. %
{} The structure itself can be arbitrarily complex because Eqs.~\eqref{eq:qm_SLRF}--\eqref{eq:qm_quadratic}
do not depend on~$N$ in any way. 

To conclude this section, let us note that the structures in question
appear to possess other conservation relations. As can be seen, e.g.,
in Fig.~\ref{fig:optical_dos}a--c, the transmission spectra contain
the same number of resonance peaks in the interval $\left[0;2\omega_{0}\right]$,
namely, nine, which equals the number of constituent layers. Bearing
a loose resemblance to the energy level splitting in solids if one
regards the layers as {}``atoms'', this was found to be a general
property of such multilayers \cite{Nano03}. However, additional restrictions
on the structures seem to be necessary, such as the outermost layers
of the structure being 1 rather than 0 (compare, e.g., Figs.~\ref{fig:optical_dos}d,~e).
This requires additional investigations and remains a subject for
further studies.

\section{Conclusions and outlook\label{sec:summary}}

To summarize, we have investigated the possible degree of modification
to transmission and dispersion properties, as well as the averaged
local DOS, in discretely inhomogeneous media. Both electromagnetic
waves propagating in a dielectric multilayer structure and a quantum
particle propagating over a stepwise, multiple-well potential barrier,
have been considered (Fig.~\ref{fig:structure}). In both cases,
certain constraints on the constituent elements of the structure {[}Eqs.~\eqref{eq:qwave}~and~\eqref{eq:qm_static}]
allow to derive the conservation relations over finite frequency or
energy regions {[}Eqs.~\eqref{eq:opt_constraint}--\eqref{eq:opt_integral_norm_0}
and~\eqref{eq:qm_constraint}, respectively]. Both relations hold
\emph{regardless} of the structure geometry (at least in the sense
of rearrangement of constituent elements) and are, in this sense,
universal, despite the fact that the spectral properties \emph{themselves}
can be strongly geometry-dependent. The quantum case appears to be
more complicated than the optical one and requires more conditions
to be fulfilled, as implied by a quadratic character of Eq.~\eqref{eq:qm_quadratic}.
The analytical results obtained have been verified by numerical calculations
(see Figs.~\ref{fig:optical_dos},~\ref{fig:quantum_dos}, and Table~\ref{tab:integral_dos}).

The results obtained contribute to understanding the physics of complex
inhomogeneous media. They can be applied in the design of heterostructures
with engineered dispersion, such as chirp compensation, pulse compression
or delay line devices. A more detailed studies of the relations obtained
would also be useful. It would be of interest to find out if, and
to what extent, the results can be applied to the case of optical
multilayers made of dispersive and/or absorptive materials, as well
as for potential barriers in the tunneling regime.

\begin{acknowledgments}
The authors are thankful to J.~P.~Dowling, H.~V.~Demir, D.~N.~Chigrin,
E.~P.~Petrov, and A.~V.~Lavrinenko for stimulating discussions,
as well as to C.~Kremers for helpful advice. Partial support from
the Basic Research Foundation of Belarus (Grant F03M-097) and the
Deutsche Forschungsgemeinschaft (Projects SPP1113 and FOR557), as
well as the support of the EC-funded projects PHOREMOST (FP6/2003/IST/2-511616)
is gratefully acknowledged.
\end{acknowledgments}

\end{document}